\documentclass[preprint]{iucr}
\journalcode{J}                       

\usepackage{amsmath}
\usepackage{amssymb} 
\usepackage[T1]{fontenc}
\usepackage{textcomp} 

\usepackage{xcolor}
\usepackage[normalem]{ulem}
\definecolor{revblue}{rgb}{0.00,0.20,0.80}
\definecolor{revred}{rgb}{0.80,0.05,0.05}
\definecolor{revgreen}{rgb}{0.00,0.45,0.15}
\newif\ifreview \reviewtrue

\usepackage{enumitem}
\let\iucrOrigCaption\caption
\usepackage[bookmarks=false,
 breaklinks=true,pdfborder={0 0 1},backref=false,colorlinks=true,hidelinks]{hyperref}
\let\caption\iucrOrigCaption

\usepackage{etoolbox}
\makeatletter
\expandafter\apptocmd\csname\string\table\endcsname
  {\ifPstyl@\else\ifVstyl@\else
     \def\hline{\noindent\kern0pt
       \hbox to 0pt{\vbox to 0pt{\hrule height .1 pt width \textwidth}\hss}%
       \\}%
   \fi\fi}
  {}{\PackageWarning{iucrhlinefix}{could not patch table hline}}
\makeatother
\usepackage{url}
\makeatletter
\g@addto@macro\UrlBreaks{\do\a\do\b\do\c\do\d\do\e\do\f\do\g\do\h\do\i\do\j
  \do\k\do\l\do\m\do\n\do\o\do\p\do\q\do\r\do\s\do\t\do\u\do\v\do\w\do\x\do\y\do\z
  \do\A\do\B\do\C\do\D\do\E\do\F\do\G\do\H\do\I\do\J\do\K\do\L\do\M\do\N\do\O
  \do\P\do\Q\do\R\do\S\do\T\do\U\do\V\do\W\do\X\do\Y\do\Z
  \do\0\do\1\do\2\do\3\do\4\do\5\do\6\do\7\do\8\do\9\do\.\do\-\do\_\do\/}
\makeatother
\renewcommand{\harvardurl}[1]{\begingroup\urlstyle{same}\url{#1}\endgroup}
\begin{document}
\title{Variable-Resolution Scattering Reveals Ensemble Properties}
\author[a,b]{Uri}{Steinitz}
\author[c]{Mathar}{Kravikass}{}{}
\author[c]{Gil}{Koren}{}{}
\cauthor[c]{Roy}{Beck}{roy@tauex.tau.ac.il}{}
\aff[a]{Soreq Nuclear Research Centre, Yavne 81800, \country{Israel}}
\aff[b]{Department of Chemical and Physical Sciences, University of Toronto Mississauga, Mississauga, Ontario L5L 1C6, \country{Canada}}
\aff[c]{The School of Physics and Astronomy, The Center for Physics and Chemistry of Living Systems, The Center for Nanoscience and Nanotechnology, and Sagol School of Neuroscience, Tel Aviv University, Tel Aviv 69978, \country{Israel}}

\maketitle

\begin{synopsis}
TENOR-SAXS recovers a nanoparticle ensemble's scattering-weighted polydispersity directly from a single two-dimensional SAXS/SANS image, by comparing two digitally applied anisotropic smearings, without requiring a model of the instrument's native resolution function.
\end{synopsis}

\begin{abstract}
Small-angle X-ray and neutron scattering (SAXS/SANS) determine the size and structure of particles and macromolecules in solution. However, the conventional low-angle determination of the radius of gyration, $R_g$, cannot separately resolve its scattering-weighted mean and the width of its distribution. Because the apparent $R_g$ includes a contribution from ensemble variance, the distribution width is required even to recover the mean $R_g$. Here, we introduce TENOR-SAXS, a method for estimating the scattering-weighted width of an $R_g$ distribution by applying two anisotropic digital smearings to a single two-dimensional scattering image. Comparing the smeared images provides curvature information unavailable from the conventional radial profile. The resulting observables require no explicit model of the native instrumental point-spread function and show only weak sensitivity over the range of native resolutions typical of SAXS measurements examined here. Analytical expressions are derived in the low-angle and narrow-distribution limits for ensembles described by a known or independently constrained single-particle form factor. For verification, we simulated scattering by particles with several form factors and ensembles with different size distributions, all agree well with the analytical theory. To leading order, the principal observable depends on the scattering-weighted mean and variance rather than on the distribution's functional form or higher moments. For simulated data corresponding to feasible synchrotron exposure times for biological samples, the distribution width, expressed as a fraction of its mean, was recovered with a typical discrepancy of approximately 0.1, while the scattering-weighted mean $R_g$ was recovered with an approximately 2\% relative discrepancy. TENOR-SAXS thus provides a framework for estimating ensemble width from existing two-dimensional scattering data without explicit instrumental desmearing and for correcting the variance contribution to the apparent radius.
\end{abstract}

\section{Introduction}
\raggedcolumns
Small-angle X-ray scattering (SAXS) is a widely used technique for characterizing nanostructures, providing ensemble-averaged information on particle size, shape, and conformational flexibility. Small-angle neutron scattering (SANS) shares similar analysis techniques, although its resolution and noise characteristics differ from those of SAXS \cite{jeffries2021primer}. Since the early developments of Guinier and co-workers \cite{guinier1955,glatter1982}, SAXS has become central in nanoscopic characterization, in particular for samples measured in solution. A relevant example includes the study of nanoscale biological substances (e.g., proteins, lipids, nucleic acids and their complexes), where it complements high-resolution crystallography and NMR by accessing macromolecular length-scale and states in solution, including flexible and heterogeneous systems \cite{svergun2003,jacques2010,mohammed2024}. In soft condensed matter and materials science, SAXS is similarly applied to polymers, nanoparticles, colloids, and porous media, enabling quantitative description of morphology and hierarchical structure \cite{feigin1987,pauw2013}.

The distribution of nano-particle sizes has fundamental effects in many cases in materials science and pharmaceuticals \cite{schoepe2007,garcia2009,pastoriza2011,danaei2018}. This motivates the development of SAXS methods to observe and characterize the polydispersity of a nanoscopic ensemble. SAXS is fundamentally limited, since the measured intensity represents an average over all scatterers in the illuminated volume. Samples such as intrinsically disordered proteins (IDPs) populate heterogeneous conformational ensembles, which may be represented in part by a distribution of radii of gyration, $P(R_g)$, rather than by a single characteristic structure \cite{ahmed2020,gomes2020,song2021,koren_intramolecular_2023}. Integrative studies further demonstrate that ensembles with similar SAXS-derived mean $R_g$ values may nevertheless differ substantially in end-to-end distances, internal-distance scaling, shape, and conformational fluctuations \cite{gomes2020}. Thus, a mean $R_g$ obtained by conventional SAXS analysis can mask structurally and functionally relevant heterogeneity and, in a polydisperse ensemble, may also be shifted by the underlying $R_g$ distribution. Approaches to recover ensemble heterogeneity include the ensemble optimization method \cite{bernado2007,tria2015}, Bayesian inference \cite{hummer2015,antonov2016,chatzimagas2023}, molecular form factor \cite{riback2017} and related modeling frameworks. Recently, Tomchuk et al. introduced a new analysis protocol to extract polydispersity directly from SAXS \cite{tomchuk2023} by extracting observable values from the data. However, the quantitative interpretation of this analysis is sensitive to the assumed functional form of the particle-size distribution, which is generally unknown, and is derived for spherical particles.

Instrumental effects provide an additional complication. Deliberate manipulation of the effective SAXS point-spread function has itself been used to extract information, by modifying the incoming beam shape and translating the detector in sub-pixel steps to retrieve an angularly super-resolved pattern \cite{gutman2020}. Beam divergence and detector resolution broaden the measured scattering profile through convolution with the point-spread function (PSF, or resolution function). This broadening mimics the effect of particle-size polydispersity, so that accurate recovery of ensemble variance from a single scattering profile requires knowledge of the PSF. However, PSF characterization is often incomplete and may vary with beamline configuration \cite{pedersen1990,vad2010,rennie2013}. Although deconvolution approaches are sometimes used in principle to correct for instrumental broadening, deconvolution itself requires accurate knowledge of the PSF and is typically ill-posed; as a result, it propagates PSF uncertainty directly into the inferred variance and can introduce systematic bias \cite{tikhonov1977,bertero1998,svergun1988,vad2011,glatter1974}.

Moment expansions of the resolution kernel have recently been used to desmear anisotropic two-dimensional scattering data \cite{huang2025,tung2025,tung2026gp,tung2026aniso}, and instrument resolution is already varied deliberately on some SANS instruments \cite{radulescu2015}; digital desmearing of measured 2D SAXS patterns dates back to \cite{leflanchec1996}, and fitting smeared models to raw data instead of desmearing was proposed even earlier \cite{sjoberg1978}. TENOR-SAXS differs in that it does not attempt to recover the desmeared intensity at all: it takes the ratio of two \emph{applied} smearings, which is only weakly sensitive to the native PSF in the tested small-smearing regime, and infers the ensemble variance directly from it.

\begin{figure}
\includegraphics[width=0.8\textwidth,trim={0cm 0cm 0 0cm},clip]{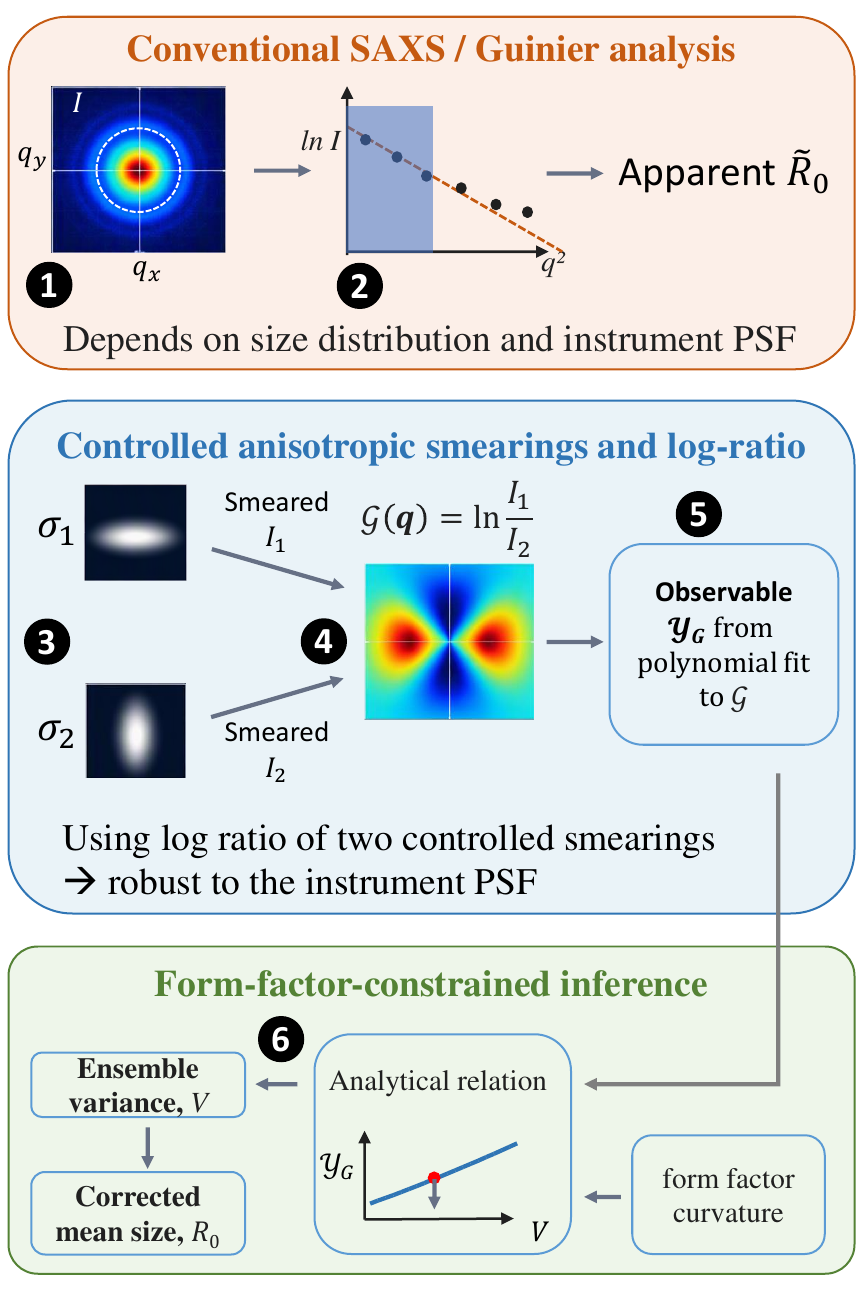}
\caption{The TENOR-SAXS workflow steps: (1) A single background-subtracted two-dimensional pattern $I(q)$; the native point-spread function of the instrument need not be known, and the dashed circle marks $q\tilde{R}_0 = 1$. (2) A straight-line Guinier fit (blue) to noisy data inside that window (shaded) returns the apparent radius of gyration, $\tilde{R}_0$. (3) The same image convolved with two anisotropic Gaussian kernels. (4) The pixelwise log-ratio of those two maps, $\mathcal{G}(q) = \ln I_1 - \ln I_2$, nearly eliminating the common instrumental smearing. (5) the radial and directional components of $\mathcal{G}$ are fitted to second-order polynomials in $q^2$. An observable $\mathcal{Y}_G$ is derived from the polynomial coefficients. (6) Knowing the form factor gives an analytic dependence $\mathcal{Y}_G(V)$, and comparison of the observable against it extracts $V$ and then $R_0 = \tilde{R}_0/\sqrt{1+V}$.\label{fig:concept}}
\end{figure}

In this work, we introduce a \textbf{T}echnique for \textbf{En}semble \textbf{O}bservation by \textbf{R}esolution variation in \textbf{SAXS} (\textbf{TENOR-SAXS}), which avoids explicit deconvolution and reduces sensitivity to the native instrumental PSF by comparing controlled changes in effective resolution. The method leverages changes in digital filtering resolution to construct observables whose dependence on the variance of $R_g$ is analytically defined and is only weakly sensitive to the shape of the examined size-distribution families when expressed in terms of scattering-weighted variance. Scattering patterns acquired at different effective resolutions, either experimentally (e.g.\ via slit or collimation adjustments) or by applying digital smearing, are compared to separate instrumental and structural contributions to profile broadening, see Fig.\,\ref{fig:concept} for a visual conceptualization of the method. In this way, the ensemble's polydispersity is characterized without requiring explicit knowledge of the PSF and without the systematic bias associated with deconvolution-based approaches.
Below, we present the theoretical formulation of the approach, validate it using analytical models, and demonstrate through simulation that ensemble variance can be recovered with a known or independently constrained monodisperse form factor family. We further evaluate the effects of experimental noise and show that signal levels typical of modern synchrotron SAXS measurements are compatible with reliable implementation.

These results establish TENOR-SAXS as a practical means of quantifying structural dispersion in solution-based ensembles, while substantially reducing native-PSF-induced bias within the small-smearing regime. Currently, the method works best when the monodisperse form factor is known, and within plausible skewness assumptions it has only a small dependence on the particle size distribution function. As such, it is applicable to ensembles of nano-scale assemblies with predetermined shapes. Future work will aim to use the approach on particles with less distinct shapes such as disordered proteins and polymers.

\section{Theory}
\subsection{Scattering off a polydisperse ensemble}
\label{sec:scattering-polydisp}
In scattering experiments (e.g., SAXS/SANS), the scattered intensity is measured as a function of the scattering vector $q = \frac{4\pi}{\lambda}\sin\theta$, where $\lambda$ is the wavelength of the X-rays or neutrons and $2\theta$ is the scattering angle. In an ideal setting of zero instrumental smearing, the measured intensity $I(q)$ reflects an average over all particles in the illuminated volume. For a system of identical particles (monodisperse), this intensity is directly related to the single particle form factor. However, as recognized by Guinier, many samples are polydisperse, meaning that structural properties such as the $R_g$ are distributed around a mean value rather than being fixed \cite{guinier1955}.

Therefore, for heterogeneous systems, the observed (normalized) scattering can be written as an ensemble average
\begin{equation}
I(q) = \big\langle F(q, R_g) \big\rangle\ ,
\label{eq:ensemble_average}
\end{equation}
where $F(q, R_g)$ is the scattering intensity from a particle of size $R_g$, and the brackets denote averaging over the distribution of $R_g$ values in the ensemble. We note here that the form factor averaging is weighted, including both the numerical abundance of each $R_g$ and the scattering strength. In the analysis presented here we use numerical abundance only, so the scattering strength is implemented by altering the distribution function. To quantify the spread of this distribution, we define its scattering weighted mean as $R_0$ and introduce a dimensionless variance
\begin{equation}
V = \frac{\operatorname{Var}(R_g)}{R_0^2},
\end{equation}
which uses, as with the mean value, the scattering strength weighted distribution. Unless otherwise stated, the variance considered here is the relative variance of the scattering-intensity-weighted $R_g$ distribution, rather than that of the particle-number distribution. Specifically, if $P(R_g)$ is the number probability density and $w$ is the scattering-weighting power, the weighted mean is \[ R_0=\langle R_g\rangle_w = \frac{\int R_g\cdot R_g^{w}P(R_g)\,dR_g} {\int R_g^wP(R_g)\,dR_g}, \] with the unweighted moments recovered by setting $w=0$. This distinction should be made explicit because the polydispersity literature commonly defines the numerical distribution width $p=\sigma_{R_g}/\langle R_g\rangle$ from the unweighted particle-number distribution \cite{tomchuk2023}. The scattered-intensity map, however, is governed exclusively by the scattering-weighted distribution. Accordingly, the variance relevant to the present analysis is $V$, defined from that weighted distribution, rather than the variance of the underlying number distribution. Because the scattered intensity of a particle scales as the square of its mass, homogeneous solid spheres of constant density and contrast enter the average with an $R_g^6$ weighting, which emphasizes larger particles. The relationship between $p$ and $V^{1/2}$ therefore depends on both the weighting power and the shape of the number distribution. Thus, for a normal number distribution restricted to positive radii, the scattering-weighted relative variance approaches a finite, weighting-dependent limit as the underlying distribution is broadened. Consequently, a large unweighted polydispersity $p$ does not necessarily imply a large scattering-weighted polydispersity $V^{1/2}$.

If the distribution is narrow ($V\ll 1$), we can expand $\ln F(q,R_g)$ around $R_0$, to leading order in $V$:
\begin{equation} \ln I(q) \approx \ln F(q,R_0) + \frac{1}{2} V R_0^2 \left[ \frac{\partial^2 \ln F}{\partial R_g^2} + \left( \frac{\partial \ln F}{\partial R_g} \right)^2 \right]_{R_g=R_0}. \label{eq:variance_expansion} 
\end{equation} 
The term linear in $R_g-R_0$ vanishes because $\langle R_g-R_0\rangle=0$. At small angles, we define the dimensionless Guinier variable $Q\equiv q^2R_0^2\ll 1$. For the Guinier form factor, the second-derivative contribution in Eq.~\eqref{eq:variance_expansion} is of order $VQ$, whereas the squared first-derivative contribution is of order $VQ^2$. Therefore, retaining terms through $\mathcal{O}(VQ)$ gives

\begin{equation}
\ln I(q) \;\approx\; \ln F(q, R_0) \,+\, \tfrac{1}{2} V R_0^2 \,\frac{\partial^2}{\partial R_g^2} \ln F(q, R_g)\Big|_{R_g = R_0}.
\label{eq:variance_correction}
\end{equation}

Converting the distribution widths reported or depicted for such systems \cite{ullman2011,Kikhney2015,ahmed2020,pauw2017} into the relative variance gives $V$ in the approximate range 0.01--0.25.
\begin{equation}
F(q,R_g) \approx I_0\exp\!\left(-\tfrac{1}{3} q^2 R_g^2\right).
\label{taylor-F}
\end{equation}
Substituting Eq.\,\eqref{taylor-F} into Eq.~\eqref{eq:ensemble_average} and retaining terms to leading order in $q^2$ yields
\begin{equation}
\ln I(q) \;\approx\; \ln I_0 - \tfrac{1}{3} q^2 R_0^2 \,(1 + V) = \ln I_0 - \tfrac{1}{3} q^2 \tilde{R}_0^2 .
\label{eq:guinier_variance}
\end{equation}

Equation~\eqref{eq:guinier_variance} shows that polydispersity manifests as an effective increase in the apparent radius of gyration when analyzed in the Guinier regime. This is the reason the scattering-weighted variance matters operationally, whatever one makes of its physical interpretation: it is precisely $V$, and not the unweighted width $p$, that appears in 
\begin{equation}
   \tilde{R}_0=R_0\sqrt{1+V}.
   \label{eq:define_R_tilde}
\end{equation}
We note that using higher orders of the Guinier analysis was done in \cite{zheng2018}, and the analysis here directly adds polydispersity aspects to it. 

Following, any Guinier-type analysis reported for a polydisperse ensemble is biased by the weighted variance, and recovering $R_0$ itself requires $V$ to be measured rather than assumed. For a monodisperse system, the slope of $\ln I (q^2)$ is proportional to $R_0^2$, the additional factor $(1+V)$ is exact at this order and the distribution shape first enters at $\mathcal{O}(q^4V^{3/2})$. Therefore, Guinier analysis \emph{per se} does not distinguish between a monodisperse sample with a larger radius of gyration and an ensemble with a distribution of radii \cite{moller1995}. Below, we will show a method to directly estimate both the variance and the mean radius of gyration.  

\subsection{Resolution effects and convolution with the instrumental point-spread function}
\label{sec:psf-smearing}

Let us consider a monodisperse sample where the measured intensity is the convolution of the intrinsic form factor \(F(\mathbf q)\) with the instrument point-spread function (PSF) \cite{guinier1955,pedersen1990,pedersen1997}:
\begin{equation}
I(\mathbf q) \;=\; R (q) \circledast F(q) \;=\; \int R(-\mathbf p)\,F(\mathbf q+\mathbf p)\,d^2p\ ,
\label{eq:conv-main}
\end{equation}

where $R(\mathbf q)$ represents the instrumental PSF, normalized to unit integral. Assuming $R$ is even, the sign of $\mathbf p$ in the convolution integral is immaterial.

In typical small-angle setups, rectangular aperture slits and anisotropic beam profiles make $R(\mathbf q)$ well approximated by an anisotropic Gaussian \cite{pedersen1990,vad2010,grabcev2007,barker1995, bergenholtz2016,li2008scatterless},

\begin{equation}
R(\mathbf q) \;=\; \frac{1}{2\pi\sigma_x\sigma_y}\,
\exp\!\Big(-\frac{q_x^2}{2\sigma_x^2}-\frac{q_y^2}{2\sigma_y^2}\Big),
\label{eq:R-gauss}
\end{equation}
with widths \(\sigma_x\) and \(\sigma_y\) along the aperture axes, assumed to be aligned with the detector.

When the PSF widths are small on the scale set by the curvature of $\ln F$, i.e.\ $\sigma^2 R_0^2\ll1$, a second-order expansion of the convolution gives the following expression for the log-intensity (see Appendix \ref{app:direct-logI} for derivation):
{
\begin{equation}
\begin{split}
\ln I(\mathbf q)\;\approx\;& \ln F(q)
\;+\;\tfrac{1}{2}\big(\sigma_x^2\,\partial_{q_x}^2+\sigma_y^2\,\partial_{q_y}^2\big)\ln F(q) \\
&\;+\;\tfrac{1}{2}\Big[\sigma_x^2\big(\partial_{q_x}\ln F( q)\big)^2+\sigma_y^2\big(\partial_{q_y}\ln F( q)\big)^2\Big]
\end{split}
\label{eq:logI-operator}
\end{equation}
}
For samples with isotropic form factors, i.e.\ $F(\mathbf q)=F(|\mathbf q|)$, either by having spherical symmetry or due to angular ensemble averaging, Eq.~\eqref{eq:logI-operator} shows that the PSF-smearing correction to the log-intensity depends on the radial derivative of $\ln F$, with anisotropic weights set by \(\sigma_x^2\) and \(\sigma_y^2\).
Equation \eqref{eq:logI-operator} also demonstrates the resemblance between the effect of PSF and the polydispersity (Eq.\,\eqref{eq:variance_correction}). This work shows how to decouple the two effects with only limited \emph{a priori} knowledge.

\subsection{Extraction of the variance from 2D measurements}
\label{sec:extraction2D}

We consider two measurements of the same sample taken with different resolutions characterized by anisotropic Gaussian widths $(\sigma_{x,1},\sigma_{y,1})$ and $(\sigma_{x,2},\sigma_{y,2})$. These PSFs may be realized either by taking two measurements with different settings, e.g., different sample to detector distances, slit sizes, or SANS-TOF wavelength resolution \cite{radulescu2015}. Alternatively, digitally smearing a single measurement using two different filters accomplishes the same effect without the need for normalization, center realignment, background subtraction and other measurement adjustment procedures.

Let $F(q)$ be the isotropic intrinsic form factor intensity (e.g.\ ideal instrument, no interparticle correlations) and $I_k(\mathbf q)$ the measured intensity for experiment $k\in\{1,2\}$, and the PSF width differences as:
\begin{equation}
\Delta\sigma_x^2=\sigma_{x,1}^2-\sigma_{x,2}^2,\qquad
\Delta\sigma_y^2=\sigma_{y,1}^2-\sigma_{y,2}^2.
\end{equation}
Using Eq.\,\eqref{eq:logI-operator}, the log-ratio of the two intensities is:
\begin{equation}
\begin{aligned}
\mathcal{G}(\mathbf q)&\;\equiv\;\ln I_1(\mathbf q)-\ln I_2(\mathbf q)
\\
&\;=\;\frac{1}{2}\,\Delta\sigma_x^2\Big[\partial_{q_x}^2\ln F+(\partial_{q_x}\ln F)^2\Big]
\\
&\;+\;\frac{1}{2}\,\Delta\sigma_y^2\Big[\partial_{q_y}^2\ln F+(\partial_{q_y}\ln F)^2\Big].
\label{eq:G_cart}
\end{aligned}
\end{equation}

Given $q=\sqrt{q_x^2+q_y^2}$ on the detector plane, and the in-plane azimuthal angle $\chi$ (i.e., $q_x=q\cos\chi, q_y=q\sin\chi$), we can substitute the Cartesian derivatives in Eq.~\eqref{eq:G_cart} to:
\begin{equation}
\begin{aligned}
\mathcal{G}(q,\chi)&\;=\;\frac{1}{2}\,\Sigma_0\Big[f'(q)+f(q)^2+\frac{f(q)}{q}\Big]
\\
&\;+\;\frac{1}{2}\,\Delta_0\Big[f'(q)+f(q)^2-\frac{f(q)}{q}\Big]\cos 2\chi,
\label{eq:G_polar_compact}
\end{aligned}
\end{equation}
where, 
\begin{equation}
\begin{aligned}
f(q)\equiv\frac{d}{dq}\ln F(q),\qquad \Sigma_0\equiv\tfrac{1}{2}(\Delta\sigma_x^2+\Delta\sigma_y^2), \\
f'(q)\equiv\frac{d}{dq}f(q),\qquad
\Delta_0\equiv\tfrac{1}{2}(\Delta\sigma_x^2-\Delta\sigma_y^2).
\end{aligned}\label{eq:define_Sigma}
\end{equation}

For compactness we define the radial functions:
\begin{equation}
\begin{aligned}
G(q)&\;=\;f'(q)+f(q)^2+\frac{f(q)}{q},
\\
M(q)&\;=\;f'(q)+f(q)^2-\frac{f(q)}{q},
\end{aligned}
\label{eq:GM_defs}
\end{equation}
and then Eq.~\eqref{eq:G_polar_compact} becomes:
\begin{equation}
\,\mathcal{G}(q,\chi)\;=\;\tfrac{1}{2}\,\Sigma_0\,G(q)\;+\;\tfrac{1}{2}\,\Delta_0\,M(q)\cos 2\chi\, .
\label{eq:G_final}
\end{equation}
Importantly, Eqs.\,\eqref{eq:GM_defs}--\eqref{eq:G_final} were derived for a monodisperse sample, so $F$ is the monodisperse form factor. When the method is applied to a polydisperse ensemble, the profile the instrument smears is the ensemble-averaged intensity $\langle F\rangle$ of Eq.\,\eqref{eq:ensemble_average} rather than $F$ itself, and $f$ is redefined accordingly as $f=\partial_q\ln\langle F\rangle$. This substitution is what carries the polydispersity into $G$ and $M$: it is the reason the coefficients of Tab.\,\ref{tab:GM-coef} depend on $V$. Moreover, the smearing expansion and the polydispersity expansion of Eq.\,\eqref{eq:variance_expansion} are formally the same operation of a second-order broadening of $\ln$ intensity that is applied in $q$ and in $R_g$ respectively. This is why a single pair of measurements can separate smearing expansion and the polydispersity expansion.

Equation~\eqref{eq:G_final}, which is fully derived in Appendix \ref{app:direct-logI}, shows that the dependence on the two experimental resolution settings is only experienced through the pre-factors $\Sigma_0$ and $\Delta_0$ representing the PSFs' isotropic and anisotropic differences, respectively. 
The azimuth-independent contribution is proportional to $G(q)$, while the twofold anisotropic contribution is proportional to $M(q)\cos 2\chi$. 
Changing the PSF widths between experiments simply rescales these contributions, while the intrinsic functions $G(q)$ and $M(q)$ contain the raw ensemble information.

If we take a monodisperse particle with a specific $R_g$, it is natural to express its form factor through a dimensionless parameter $u=(qR_g)^2$. Following, we can expand the form factor logarithm around the origin:
\begin{equation}
\ln F_{R_g}(q) = \phi(u) = \kappa\,u + \tfrac{1}{2}\phi''\,u^2 + \mathcal{O}(u^3)
\end{equation}
with
\begin{equation}
\kappa=\partial_u\phi(0),\;\; \phi''=\partial^2\phi(0)/\partial u^2\;.
\label{eq:phi_series u}
\end{equation}
At low scattering angles, the Guinier rule applies \cite{guinier1955}, and $\kappa=-\tfrac{1}{3}$. Note that we limit our current discussion to the second order in $u$. In Tab.\,\ref{tab:Typical-values-for} we show some examples for typical form factors but one may readily derive the relevant $\phi''$ from any specific form factor, by taking the Maclaurin series of $\ln{F}$ with respect to $u$. For the canonical form factors of Tab.\,\ref{tab:Typical-values-for}, $|\phi''|\le0.06$, so the terms of order $(\phi'')^2\lesssim4\times10^{-3}$ neglected below are small; this is the sense in which $\phi''$ is treated as a small parameter throughout.

For a collection of particles sharing an identical form factor functional dependence (i.e.\ a fixed $\phi''$), but exhibiting a distribution of radii of gyration $P(R_g)$, the polydisperse scattering intensity $I(q)$ can be expressed using a cumulant expansion. This method was originally introduced by Thiele \cite{thiele1889} and developed in modern form by Cram\'er \cite{cramer1946}, and has previously been applied to size polydispersity in dynamic light scattering \cite{koppel1972} (see Appendix~\ref{app:cumulant}).
\begin{table}
\caption{Canonical monodisperse form factors and their relevant $\kappa,\phi''$ coefficients defined
by the dimensionless derivatives in Eq.\,\eqref{eq:phi_series u} and obtained by Maclaurin expansion of $\ln F$ in $u$. The scattering strength radius weighting power ($\propto\text{mass}^2$) is indicated as well. For the Gaussian chain the listed power of $0$ corresponds to an ensemble of conformations at fixed contour length (constant mass), as used in Sec.\,\ref{sec:sim_valid}; a Gaussian chain with a distribution of contour lengths would instead carry a weighting power of $4$, since $R_g^2=Ll_p/3$. Sources: spherical shell, \citeasnoun{guinier1955}; solid sphere, \citeasnoun{rayleigh1910}; thin rod, \citeasnoun{pedersen1997}; Gaussian chain, \citeasnoun{debye1947}.}
\label{tab:Typical-values-for}
\begin{tabular}{lccc}
Monodisperse form factor & $\kappa$ & $\phi''$ & Weighting power \\
\hline
Spherical shell & $-1/3$ & $-1/45$   & 4 \\
Solid sphere    & $-1/3$ & $-1/63$   & 6 \\
Thin rod        & $-1/3$ & $+11/225$ & 2 \\
Gaussian chain  & $-1/3$ & $+1/18$   & 0 \\
\end{tabular}
\end{table}
The expansion reaches a polynomial form in $q^2$. It is convenient to write it directly for the measured map of Eq.\,\eqref{eq:G_final}, since that is what is fitted to the data:
\begin{equation}
\mathcal{G}(q,\chi)=\sum_{n\ge0} g_n (q^2)^n\;+\;\cos 2\chi\sum_{n\ge0} m_n (q^2)^n,
\label{eq:GM-poly}
\end{equation}
so that $g_n=\tfrac{1}{2}\Sigma_0\,[G]_n$ and $m_n=\tfrac{1}{2}\Delta_0\,[M]_n$, where $[G]_n$ and $[M]_n$ are the coefficients of the intrinsic functions defined in Eq.\,\eqref{eq:GM_defs}. The coefficients $g_n$ and $m_n$ therefore depend on $R_0$, $V$ and $\phi''$ and, through the prefactors, on the two applied PSFs. Approximate derivations for the first three coefficients are presented in Tab.\,\ref{tab:GM-coef}. 
\begin{table}
\caption{Analytical expressions for the polynomial coefficients of the measured log-ratio map $\mathcal{G}$, Eq.\,\eqref{eq:GM-poly}, up to first order in $V$ and $\phi''$. Omitted $\phi^{(3)}$ contribute only to $g_2$ and $m_2$ terms.}
\label{tab:GM-coef}
\begin{tabularx}{\linewidth}{cXX}
$n$ & Isotropic coefficients ($g_n$) & $\cos2\chi$ coefficients ($m_n$) \\
\hline
0 & $-R_0^2\cdot \tfrac23\Sigma_0(1+V) = -\tfrac23\Sigma_0\,\tilde{R}_0^2$ & $0$ \\[4pt]
1 & $R_0^4 \cdot\tfrac29 \Sigma_0$ & $R_0^4\cdot\tfrac29 \Delta_0$\\
~ & $\left(1+18\phi'' + \!10V+ 108V\phi''\right)$ & $\left(1+9\phi'' + \!6V+ 54V\phi''\right)$ \\[4pt]
2 & $-R_0^6\!\cdot\,\tfrac43\Sigma_0\left(\phi'' + \!\tfrac49V+ 25V\phi''\right)$ & $-R_0^6\!\cdot\, \tfrac43\Delta_0\left(\phi'' + \!\tfrac49V+ 19V\phi''\right)$ \\
\end{tabularx}
\end{table}

\subsection{2D extraction of the functional form}
Equation~\eqref{eq:G_final} allows an effective separation of variables once we have a 2D log-ratio of two measurements. It is possible to fit (e.g., by the least-squares method) the functional form of the map of $\mathcal{G}$ to an azimuth-independent part and a twofold anisotropic component. This fitting scheme is data-efficient as it avoids azimuthal binning and draws information directly from every raw pixel \cite{pauw2010,paul2013}.

Fitting the map to the polynomial form of Eq.\,\eqref{eq:GM-poly} returns coefficients that still carry the scaling factors $\tfrac{1}{2}\Sigma_0$ and $\tfrac{1}{2}\Delta_0$, which depend on the two smearing PSFs used. However, it is possible to extract scale-independent information using \emph{ratios} of coefficients from the same family, such as $g_1/g_0$ or $m_2/m_1$, in which these factors cancel identically. 

Using Eq.\,\eqref{eq:guinier_variance} and a Guinier analysis we can estimate $\tilde{R}_0^2=R_0^2\,(1+V)$ and define \textit{dimensionless observables} which originated directly from the data, such as:
\begin{equation}
\label{eq:observables-old}
\mathcal{J}_G \equiv \frac{g_1/g_0}{\tilde{R}_0^2}\approx\frac{\,1+18\phi'' + 10V\!+108V\phi''}{-3\,(1+V)^2}
\end{equation}

\begin{equation}
\mathcal{J}_{G2} \equiv \frac{g_2/g_1}{\tilde{R}_0^2}\approx\frac{18\phi''+8V+450\,V\phi''}{\,-3(1+V)(1+18\phi'' + 10V+108\,V\phi'')}
\end{equation}

\begin{equation}
\mathcal{J}_M \equiv \frac{m_2/m_1}{\tilde{R}_0^2}
\approx \frac{18\phi'' + 8 V +\! 342\,V\phi''}
{-3(1+V)\left(1+9\phi'' + 6V\!+54\,V\phi''\right)}.\label{eq:R_m-observe-old}
\end{equation}

Equations \eqref{eq:observables-old}-\eqref{eq:R_m-observe-old} present three dimensionless observables of the two unknowns $V$ and $\phi''$, from which $R_0$, the PSFs, and the beams' strengths were eliminated. Below, we will show how to extract from these observables the value of $V$ when $\phi''$ is known. We note that instead of dividing by the observed radius of gyration squared ($\tilde{R}_0^2$), division by $g_0$ provides expressions which are $R_0$ independent:
\begin{equation}
\label{eq:observables}\mathcal{Y}_{G} \equiv \frac{\Sigma_0}{2}\frac{g_1}{g_0^2}\approx\frac{\,1+18\phi'' +10V\!+108V\phi''}{4\,(1+V)^2}
\end{equation}

\begin{equation}
\mathcal{Y}_{G2} \equiv \frac{\Sigma_0}{2}\frac{g_2}{g_1g_0}\approx\frac{18\phi''+8V+450V\phi''}{\,4(1+V)(1+18\phi'' + 10V+108V\phi'')}
\end{equation}

\begin{equation}
\mathcal{Y}_M \equiv \frac{\Sigma_0}{2}\frac{m_2}{m_1g_0}
\approx \frac{18\phi'' + 8V+ 342V\phi''}
{4(1+V)\left(1+9\phi'' + 6V\!+54V\phi''\right)}.\label{eq:R_m-observe}
\end{equation}

Unlike the $\mathcal{J}$ observables, the $\mathcal{Y}$ observables require the applied PSF difference $\Sigma_0$ to be known explicitly to cancel the $\tfrac{1}{2}\Sigma_0$ scale of the fitted coefficients. Nonetheless, if the secondary PSF is applied digitally to a single recording, the exact values of these differences are known. Therefore, we have a set of observables that are uniquely correlated to $V$, $\phi''$ and the digitally applied secondary PSF and do not require the extraction of the radius of gyration by Guinier analysis. We will show below how the PSF degree of freedom allows us to choose a secondary PSF to optimize observable extraction \textit{vis-\`a-vis} the measurement noise. For the different sensitivity analyses and comparison to theory, we use alternatively the $\mathcal{J}$ and $\mathcal{Y}$ observables, having relative advantages per case, but being related by a constant factor of $-4/3$.

The cumulant expansion method allows us to expand the application of the Guinier analysis, in which one fits the log-intensity versus $q^2$ curve to a polynomial form. Using cumulants, we derive the expression (see more detail in Eq.\,\eqref{eq:lnI_dimless_app} in Appendix \ref{app:cumulant}): 

\begin{equation}
\begin{aligned}
\ln I(q)
=&\; \ln I_0 - \tfrac13\,(1+V)\,Q
\\
&\;+\left[\tfrac29 V\;+\;\tfrac12\phi''\,
\;+\;\,3\,V\phi''\right]\,Q^2
\;+\; \mathcal{O}(Q^3).
\end{aligned}
\label{eq:lnI_dimless_}
\end{equation}
where $Q=(qR_0)^2$ (valid within the Guinier regime, at $Q<1$). In the monodisperse case ($V=0$) the equation reduces to Eq.\,\eqref{eq:phi_series u}. By fitting the Guinier curve to a quadratic polynomial, $\ln I(q)\approx t_0+t_1q^2+t_2 q^4 $, we may now define a new observable 
\begin{equation}
\mathcal{Y}_T\equiv \frac{t_2}{t_1^2} \approx\frac {9\phi''+4V +54V\phi''}{2(1+V)^2}\label{Y_T definition}
\end{equation}
This observable does not eliminate the instrumental PSF, but its functional form is independent of $R_0$. The two are not independent \emph{as calibration curves}: $\mathcal{Y}_G-\mathcal{Y}_T=(1+2V)/[4(1+V)^2]=\tfrac{1}{4}-\tfrac{1}{4}V^2+\mathcal{O}(V^3)$, independent of $\phi''$. They are, however, independent \emph{measurements}, since $\mathcal{Y}_T$ retains the native-PSF contribution that the log-ratio observables cancel. This observable is easy to obtain from the measurements and seems less susceptible to noise. The approach of extracting observables from experimental data and comparing them to results of forward-simulation for characterizing polydispersity resembles past works \cite{beaucage2004,tomchuk2023}, but requires fewer \textit{a priori} assumptions regarding the form factor and the ensemble's size distribution shape.

\section{Numerical validation\label{sec:sim_valid}}
\subsection{Noise-free simulation}\label{sec:noise_free_sim}
To validate our theory, we constructed simulations of the scattering from ensembles of particles. Each ensemble was prepared populating particles of the desired probability distribution function of the radius of gyration including its weighted mean ($R_0$) and variance ($V$). Each particle had a given monodisperse form factor, from which the simulation calculates its scattered intensity profile. The contributions from the ensemble's members are then added to a total intensity map. The simulation uses a realistic synchrotron beam setup (wavelength, distances, aperture sizes and detector resolution, see Tab.\ \ref{tab:beam_simul_param} in Appendix \ref{sensitivityAnalisys}). The simulation and analysis tools are deposited online \cite{TENORSAXS_GitHub}.

Practically, for each pixel we calculated the contribution of the scattering from each of the ensemble's members according to its radius of gyration and the form factor and summed over all particles with a scattering-strength weighted as a power of the radius. Pixels at which the quartic term of the expanded log form factor exceeds the quadratic term are discarded, since the second-order expansion is no longer valid there; depending on $R_0$ and the form factor this can remove most of the detector area and expedite the analysis. To achieve this we performed a Guinier analysis to obtain the apparent radius of gyration ($ \tilde{R}_0 $), as discussed in Sec.\,\ref{sec:scattering-polydisp}. This is mainly needed for assessing the Guinier regime and to perform the analysis only on pixels within it.
\begin{figure}
\includegraphics[width=0.35\paperwidth]{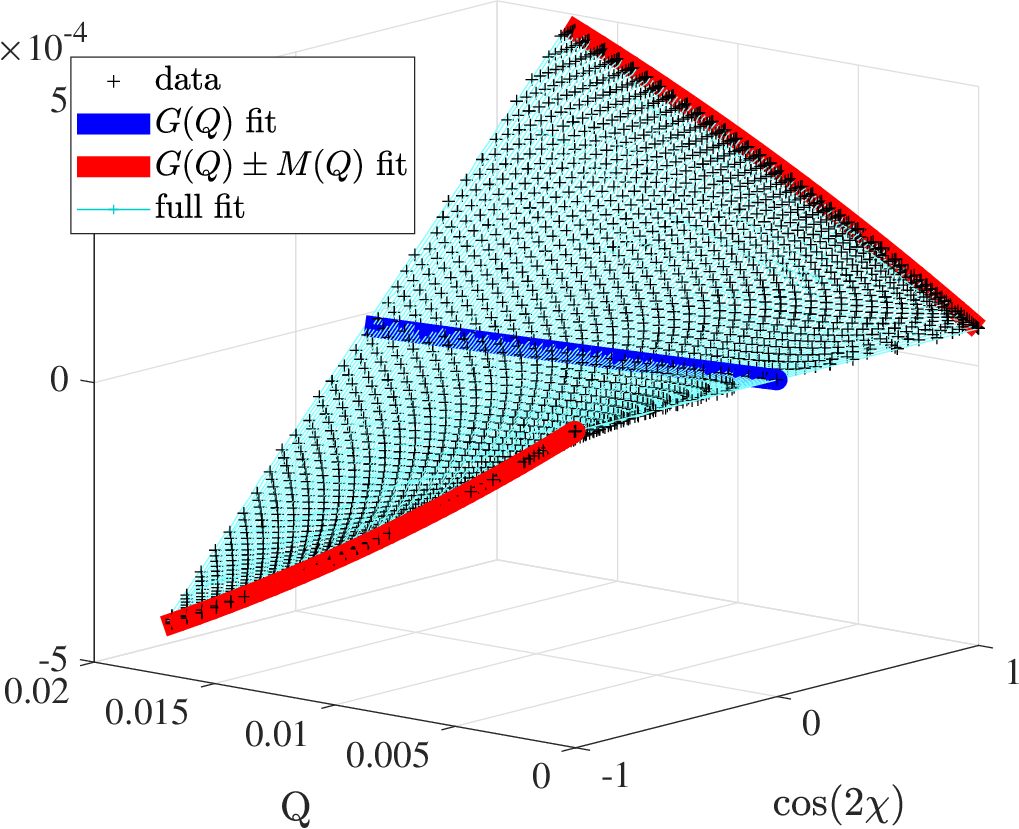}
\caption{Gaussian-chain's form factor simulation of the difference of log-intensity of a scattered intensity
after applying two different PSFs, plotted in azimuthal and radial axes ($\cos2\chi$ and $Q=q^{2}R_{0}^{2}$). The numerical fitting obtains a good
approximation and makes it easy to extract the functions $G$ and
$M$. The profile at $\cos2\chi=0$ (blue) provides $\tfrac{1}{2}\Sigma_0 G(Q)$, while the profiles at $\cos2\chi=\pm1$ (red) give $\tfrac{1}{2}\Sigma_0 G(Q)\pm\tfrac{1}{2}\Delta_0 M(Q)$. }
\label{fig:2d m_g fit}
\end{figure}
Next, we smeared the scattering maps by independently performing convolutions with anisotropic Gaussians (with different widths along the axes, $\sigma_x$ and $\sigma_y$). By comparing two such maps we obtained the log-ratio for each pixel, which may be represented in polar coordinates, $\mathcal{G}(q,\chi)$. The following step uses the least-squares method weighted by each pixel's intensity to fit $\mathcal{G}(q,\chi)$ to the radial and $\cos2\chi$ dependence (see Eq.\,\eqref{eq:G_final}). This procedure allowed us to obtain $G(q)$ and $M(q)$ as 2\textsuperscript{nd}-order polynomials in $q^2$, as demonstrated in Fig.\,\ref{fig:2d m_g fit}.  

Finally, the observables $\mathcal{Y}_G(V,\phi'')$, $\mathcal{Y}_{G2}(V,\phi'')$ and $\mathcal{Y}_M(V,\phi'')$ are formed from the fitted coefficients $g_n,m_n$ of the measured map $\mathcal{G}$ itself (Eq.\,\eqref{eq:GM-poly}), together with the known smearing widths through $\Sigma_0$ (Eqs.\,\eqref{eq:observables}-\eqref{eq:R_m-observe}).

\begin{figure}
\includegraphics[width=0.9\textwidth]{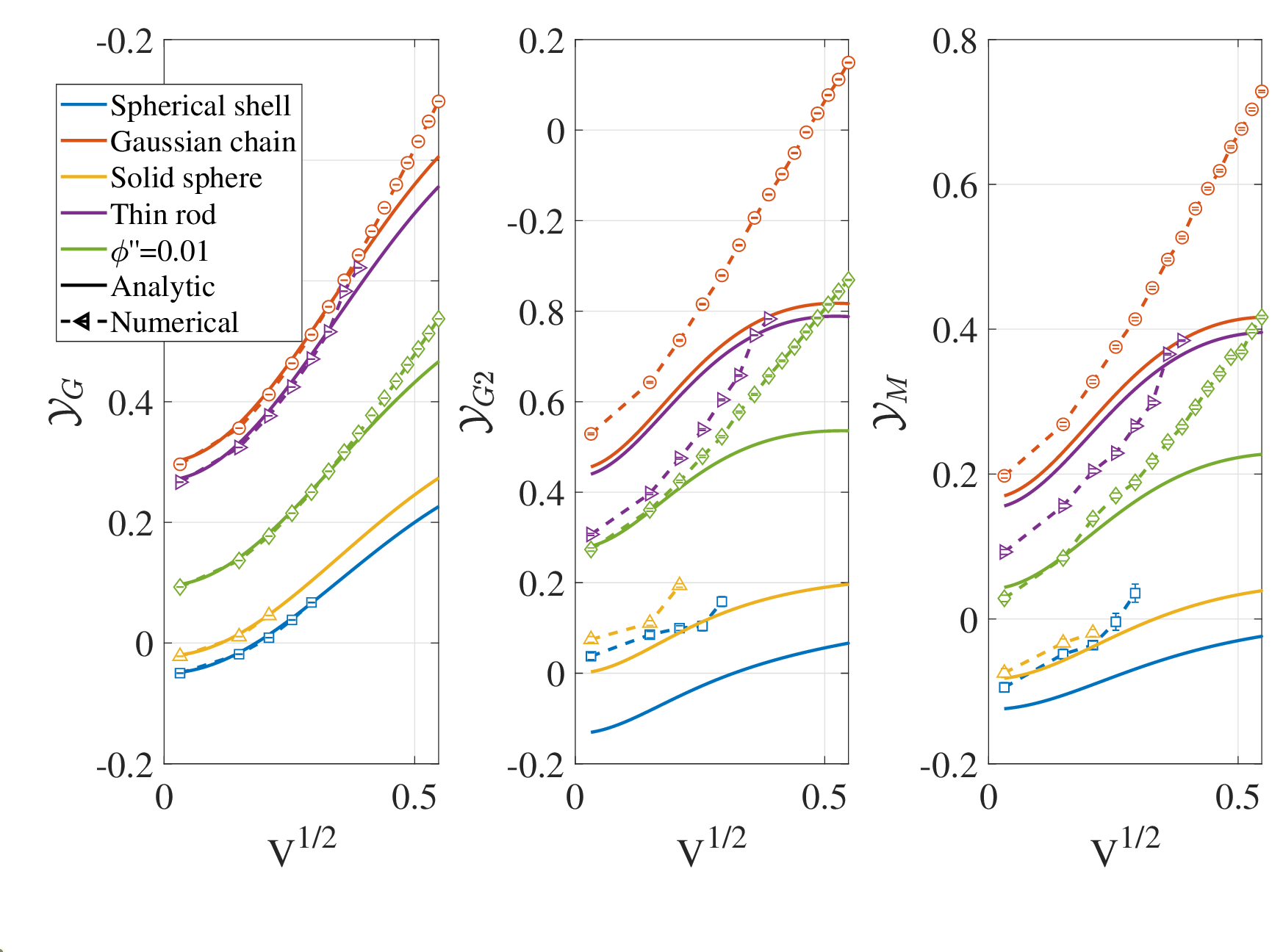}\caption{Observables obtained from 2D simulation results (dashed lines) and the theoretical first-order ratio from Tab.\ \ref{tab:GM-coef} (solid lines), plotted for a range of variances ($V$) and different form factors. Observables of $\mathcal{Y}_G$, $\mathcal{Y}_{G2}$ and $\mathcal{Y}_M$ are plotted in the left, center and right panels, respectively. In these simulations the normally distributed ensemble had an average $R_0=5\,\mathrm{nm}$, in a typical Diamond B21 beamline setup, without noise. The theoretical curves are based on the Guinier curvature $\phi''$ only, so for the exact form factors, which have higher-order terms, the theory-simulation disparity is visible for the observables based on high-order $Q$ polynomial fits- $\mathcal{Y}_{G2}$ and $\mathcal{Y}_M$.}
\label{fig:theory-sim-disparity}
\end{figure}

For the validation of our analytical derivation we performed simulations over a range of $V\in[0,0.3]$ with either exact form factors or using Guinier curvatures $\phi''\in[-0.02,0.06]$. Figure \ref{fig:theory-sim-disparity} compares the calculation results to the analytical theory, for the three observables $\mathcal{Y}$. We present the exact form factors of Gaussian chains, thin rods, spherical shells and solid spheres, along with one with an intermediate $\phi''=0.01$. The simulation of the $\mathcal{Y}_G$ observable fits the theory in these cases over a range of variances, reaffirming our theoretical derivation and the extraction method. For the form factors used, the adequate scattering mass weighting was applied. 

Some distribution families (e.g., normal distribution) have an intrinsic upper bound on their power-weighted relative variance. This limit is intrinsic rather than a consequence of the finite number of ensemble members used in the simulation. Writing the weighting as $R_g^{\,w}$, the weighted relative variance is a ratio of number-weighted moments,
\begin{equation}
V=\frac{\langle R_g^{\,w+2}\rangle\,\langle R_g^{\,w}\rangle}{\langle R_g^{\,w+1}\rangle^{2}}-1\ ,
\label{eq:V_moments}
\end{equation}
which for solid spheres ($w=6$) reads $V=\langle R_g^{8}\rangle\langle R_g^{6}\rangle/\langle R_g^{7}\rangle^{2}-1$. For a normal number distribution retaining only positive $R_g$, this ratio saturates as the distribution is broadened, approaching the half-Gaussian limit $V_{\max}=\Gamma(\tfrac{w+3}{2})\Gamma(\tfrac{w+1}{2})/\Gamma(\tfrac{w+2}{2})^{2}-1$, that is $0.178$, $0.104$ and $0.074$ for thin rods, spherical shells and solid spheres respectively, regardless of the number of ensemble members. A log-normal ensemble is not restricted in this way and its moments let $V$ become unlimited for any weighting power.

However, the $\mathcal{Y}_{G2}$ and $\mathcal{Y}_M$ observables demonstrate a distinct disparity from the analytical formulas when using exact form factor functions due to higher order curvatures ($\phi^{(3)}$, etc.) which we neglected in our analytical derivation. The $\mathcal{Y}_{G2}$ and $\mathcal{Y}_M$ observables use higher order polynomial coefficients and therefore are more sensitive to the exact form factors as well as being less accurate for higher $V$s. Nevertheless, Fig.\,\ref{fig:theory-sim-disparity} demonstrates that the actual observables' dependence on the variance is 'well-behaved' in all the examined cases. This means that a single-valued calibration curve may be extracted if the form factor is known. The $\mathcal{Y}_{G}$ observable seems independent of the high order curvature, which makes it a more practical observable to use.

Previous works suggested polydispersity-related observables which showed a dependence on the variance as well as the distribution function \cite{tomchuk2023}. A prominent reason for the sensitivity to the distribution shape is that the scattering strength of a particle is weighted by the square of its mass, or $R_g^6$ in the case of solid spheres. Thus, a number distribution will have an observed scattering which fits a weighted mean and variance. To the retained order, the observable depends on the scattering-weighted mean and variance rather than on an assumed parametric distribution. Residual dependence on distribution shape enters through neglected higher moments and is small for the distribution families and parameter ranges examined. We found that this holds quite well in practice. 

\begin{figure}
\includegraphics[width=0.95\textwidth]{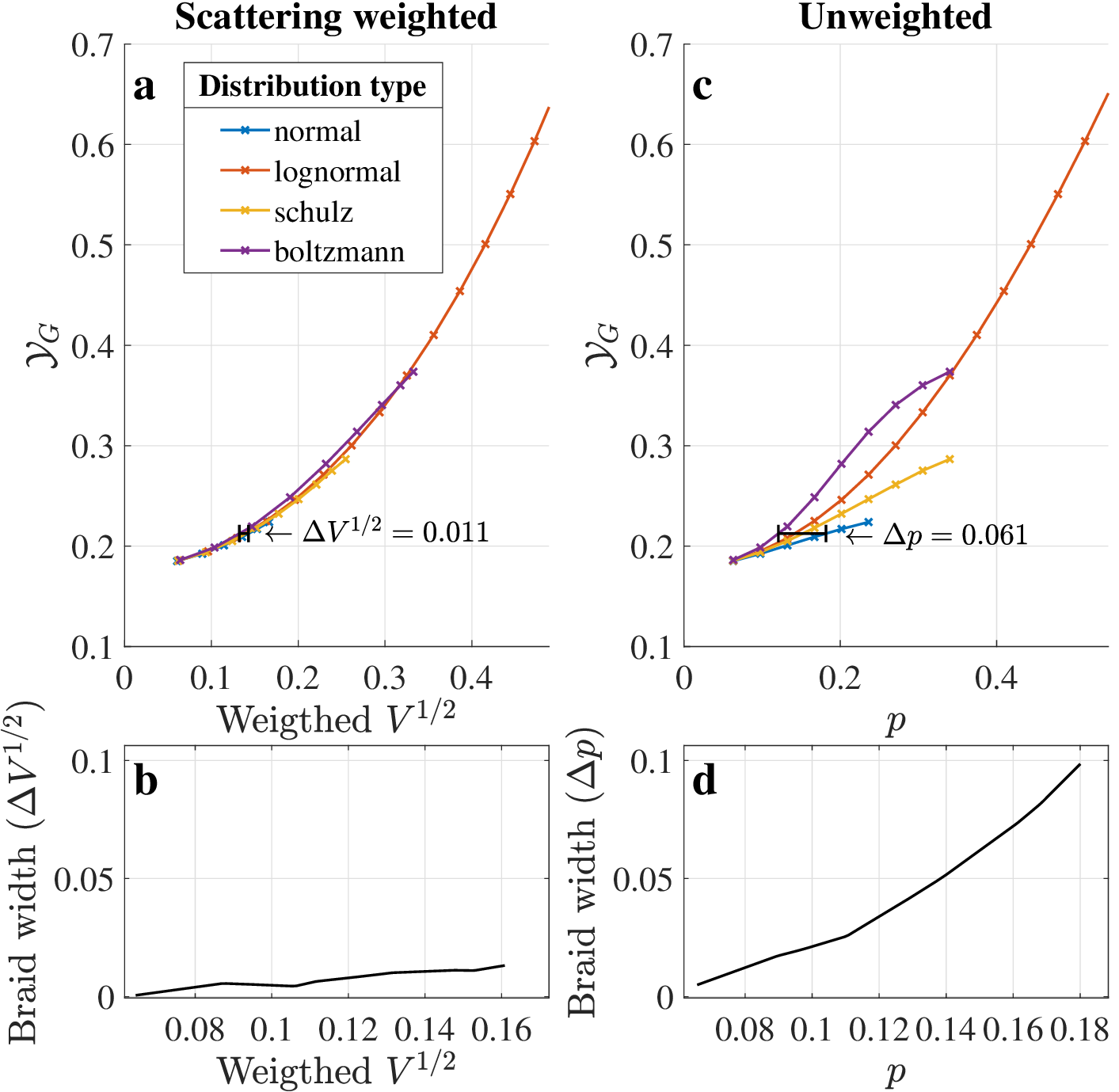}
\caption{The sensitivity to distribution type, in weighted and unweighted cases. (a) Dependence of the $\mathcal{Y}_G$ observable on the weighted polydispersity ($V^{1/2}$) for different ensemble distribution functions (solid sphere's form factor, $R_0=3\,\mathrm{nm}$, scattering-weight $\propto R_g^6$). The horizontal span of the curves' braid indicates the uncertainty of $V^{1/2}$ extraction on the distribution function (see annotated example). (b): The span of the weighted $V^{1/2}$ extraction as a function of the average $V^{1/2}$. (c)-(d) present the same data scaled by the unweighted standard deviation $p$ extraction.}
\label{fig:p_distribution_braid}
\end{figure}

Next, we simulated scattering by ensembles of solid spheres, with different variances and different distribution functions (such as normal, log-normal, uniform, triangular, Boltzmann and Schulz distributions, see \cite{tomchuk2023}), and compared the calculated $\mathcal{Y}_G$ observable (Fig.\ \ref{fig:p_distribution_braid}). With an unknown distribution function, the different distribution functions produce nearly the same observable at a given weighted-variance and therefore the uncertainty of the weighted variance extraction is very small compared to the sensitivity to other parameters, as will be shortly discussed. For the distribution families and parameter ranges examined here, scattering weighting commonly reduces the relative variance because the weakly scattering small-particle population contributes little to the measured intensity. This ordering is not universal as demonstrated by evaluating the unweighted variance of the same simulations (Fig.\ \ref{fig:p_distribution_braid}B). This discrepancy is inherent to SAXS/SANS, as adding small radii populations increases the unweighted number density variance, but their weighted scattering strength is negligible.
We extended our analysis to examine the dependence on distribution skewness; notably, \(\mathcal{Y}_{G}\) exhibits high resilience even against highly skewed distributions. The other observables, using higher polynomial orders, are more susceptible to the change in the third distribution moment. More details are presented in App.\ \ref{sensitivityAnalisys}.

We tested the effect of different parameters on the observables within the approximation assumptions (Appendix \ref{sensitivityAnalisys}). Importantly, we found that the instrumental PSF size hardly affects the observables thus demonstrating our ability to decouple PSF smearing from the population variance effect. Our evaluation included different beam shapes and sizes, spanning from below typical synchrotron beam sizes to above laboratory X-ray setups. The extracted observables are also practically independent of the average radius of gyration used in the simulated ensembles or the ensemble's size (how many different particles are used for the simulation). To leading orders, they are only correlated to $V$ and the form factor.
\subsection{Digital post-PSF filtering considerations}
Having produced the direct dependence of the ensemble properties on observables produced from the log-ratio of measured intensities with two anisotropic PSFs, we may sketch a recipe for using the method for practical application.
In order to obtain two different PSFs, one may use several methods, such as repeating the measurement with different incoming beam sizes or different sample-to-detector distances. However, a simpler route would be to take a 2D measurement and apply a digital filter, as we used in the simulations. This method is more practical for several reasons. Notably, it eliminates the need for multiple costly experiments, reduces concerns about sample stability, and avoids dependencies on fixed beam centers, slit settings, repeated buffer subtractions, normalization or other system parameters that may vary between measurements.

\subsection{Noisy simulation analysis}
Real measurements are noisy and depend on different parameters such as measurement time, source and sample stability, and more. Therefore, we do not know \textit{a priori} which smearing would be best for each observable fit. However, it is straightforward to grade the quality of the $G$ and $M$ polynomial fits (Eq.\,\eqref{eq:GM-poly}) for different PSF sets by comparing the standard errors of the resulting observables (Eqs.\,\eqref{eq:observables}-\eqref{eq:R_m-observe}). Thus, by scanning the smearing sizes, it is possible to find the best estimate of the observable available from the data.

The next step is the extraction of the ensemble parameters by comparing the $\mathcal{Y}_G$ observable to the analytical formula, and interpolating to find the best $V$ that supports the observable and the $\phi''$ values. For the observables $\mathcal{Y}_{G2}/\mathcal{Y}_{M}$, due to the lower accuracy of the first order analytic formulae but the low sensitivity to most parameters (as discussed earlier and in Appendix \ref{sensitivityAnalisys}), a comparison to a calculated calibration curve is a better route. 
Extraction of the form factor from the combination of different observables is possible in theory, but in practice requires extremely low noise levels. All variance and $R_g$ performance figures reported below were obtained from $\mathcal{Y}_G$ alone. 

\section{TENOR-SAXS analysis procedure and results}
Solution SAXS analysis is typically performed on azimuthally averaged one-dimensional datasets, which discard most information about anisotropic instrumental broadening. Here, we present a procedure for estimating the scattering-weighted mean $R_g$ and the scattering-weighted relative variance $V$ when the underlying monodisperse form factor is known or can be reliably approximated. Uniquely, TENOR-SAXS analysis is PSF-robust and performs well at feasible photon counts. Moreover, its sensitivity to the assumed particle-size distribution is small. Below we list the key steps to perform TENOR-SAXS analysis (see also, Fig. \ref{fig:concept}).
\subsection{TENOR-SAXS procedure}
\begin{enumerate}
\item Obtain a 2D SAXS or SANS measurement. After removing the background intensity, the data should have the intensity for the 2D scattering angle $I(\mathbf{q})$. Note that background measurement removal is typically done in radially averaged (1D) data, and here the 2D process should be done with care, for example, by projecting the averaged 1D background data back to 2D and then subtracting it.
\item Perform a preliminary low-$q$ analysis to obtain the apparent radius $\tilde R_0$. As a default, restrict the TENOR-SAXS fit to pixels satisfying $q\tilde R_0<1$. Pixels whose convolution neighborhood extends beyond this range or beyond the valid detector area should also be excluded. The basis for this practical fitting range is discussed below.
\item Choose digital filters to find well fitted observable values:
\begin{enumerate}

\item \textbf{Create two digital kernels} (typically Gaussian) with sizes $\sigma_{x_{1,2}}$ and $\sigma_{y_{1,2}}$ ('PSF quartet') which are well within the Guinier regime ($\sigma \tilde{R}_0\ll1$) and \textbf{calculate $\Sigma_0$} using Eq.\,\eqref{eq:define_Sigma}. Apply the digital filters by convolving the image with each digital kernel and \textbf{obtain two intensity maps}, each smeared differently. As described above, the difference between the two PSFs is important as well, so typically two PSF sizes differ by tens of percent. 

\item \textbf{Take the logarithm of the ratio of the two maps}, and perform a weighted \textbf{fit of $\mathcal{G}(q,\chi)$} to the form of Eq.\,\eqref{eq:G_final} and to \textbf{the 2\textsuperscript{nd} order polynomials $G(q)$ and $M(q)$}. Pixels were weighted approximately by intensity to reflect the leading count dependence of relative photon noise; this weighting does not account for correlations introduced by filtering. Unless stated otherwise, $G(q)$ and $M(q)$ are fitted as 2\textsuperscript{nd}-order polynomials in $q^2$. 

For the noise-free sensitivity analyses of Appendix\,\ref{sensitivityAnalisys}, a 3\textsuperscript{rd}-order fit was used, which is better conditioned in the absence of noise; the cubic term is strongly noise-sensitive and was therefore omitted in the noisy benchmark and in the recommended protocol. Fitting a cubic to data generated by a quadratic model shifts the recovered 2\textsuperscript{nd}-order coefficient by a fit-range-dependent amount, so $\mathcal{Y}_{G2}$ and $\mathcal{Y}_{M}$ must be compared with a calibration curve computed at the same polynomial order; $\mathcal{Y}_G$ is less sensitive to polynomial order because it uses only $g_0$ and $g_1$, but the fitted values may still depend weakly on model order and fit window.

\item Obtain the \textbf{$\mathcal{Y}$ observables} using Eqs.\,\eqref{eq:observables}-\eqref{eq:R_m-observe} from the polynomial coefficients of $G(q)$ and $M(q)$, and the difference between the filters $\Sigma_0$. 
The confidence interval of the fit, together with the known theoretical derivative $\partial\mathcal{Y}_G/\partial V$, indicates whether the accuracy is sufficient or a different quartet should be tried. In practice, this search is short. The kernels must be wide enough to average over many pixels yet stay within the Guinier regime, and those two requirements between them leave little room: a serviceable starting point is $\max(\sigma)\cdot\tilde{R}_0\approx1/4$, with the second kernel narrower by a few tens of percent. We have found that for the tested cases, a quartet chosen this way rarely needs more than a few refinements.
\end{enumerate}

\item \textbf{Find $V$ for each observable $\mathcal{Y}$ and calculate $R_0$}. For $\mathcal{Y}_G$, interpolate the analytical formula in Eq.\,\eqref{eq:observables} to find the value of $V$ that fulfills it. $R_0$ can then be calculated from Eq.\,\eqref{eq:define_R_tilde}. If needed, the other observables may be used by comparison to a calibration curve computed by simulating the scattered intensity of an ensemble with the known form factor, the measured $R_g$ (i.e., $\tilde{R}_0$), the intensity weighting power and different $V$ values, filtering and fitting to $G$, $M$ and extracting the observables as detailed above.

\item One may \textbf{scan alternative filters to evaluate the sensitivity} of the observables and repeat the above-mentioned steps. Typically, the best smearing kernel may be predetermined, tending toward large areas (over which the integration reduces noise), but low enough so that there are enough pixels left that are not affected by data from outside the Guinier regime. In the predetermined case, one may limit the scan to just a few cases in the vicinity of the chosen quartet.
\end{enumerate}
\subsection{Practical limitations}
The practical application of TENOR-SAXS is restricted to a low-$q$ fitting interval. Two conditions jointly determine the usable range: the upper bound imposed by the low-$q$ expansion and the exclusion of pixels whose convolution neighborhood extends beyond either this range or the valid detector area. The relative importance of these conditions depends on the smearing kernels and detector geometry. In the present implementation, we use $q\tilde R_0 < 1$ as the nominal upper bound, where $\tilde R_0$ is the apparent radius obtained from the preliminary low-$q$ analysis. Although the analytical expressions are asymptotic expansions about $Q=q^2R_0^2=0$, the suitability of this finite fitting interval was established a posteriori by comparison with numerical calculations using full form factors. For the form factors and distributions examined, this choice retains sufficient pixels for noise-tolerant fitting while keeping the systematic discrepancy in the recovered ensemble width small. The bound should therefore be regarded as a numerically validated bias or variance compromise, rather than as a sharp asymptotic threshold. For form factors that depart more rapidly from their low-$q$ expansion, a smaller upper bound or a calibration based on the full form factor may be required.
\subsection{Results}
We performed numerous simulations and scanned the applicable regime using the Diamond B21 beamline instrument configuration (see Tab.\,\ref{tab:beam_simul_param} in Appendix \,\ref{sensitivityAnalisys}) spanning $-0.02<\phi''<+0.06$, $0<V<0.3$, $2.5~\text{nm} < R_0 < 7~\text{nm}$ and with Gaussian (normal) probability distribution function. Applying the TENOR-SAXS method provided the sensitivity to different simulations and ensemble parameters, as well as to the TENOR-SAXS specifications, showing that the ground truth (noise-free) observables are mostly insensitive to them (Appendix \ref{sensitivityAnalisys}). Notably, lowering $R_0$ mainly affects the results through increasing the number of pixels within the Guinier regime, thus the 2D fit is based on additional points, reducing the fit error. 

We have performed the TENOR-SAXS protocol with simulated data with different noise levels. The noise model was simulated as Gaussian with a standard deviation equal to the square root of the number of photons per pixel \cite{gutman2020}. For this simulation we employed a normal radius distribution function of Gaussian chain's form factor \cite{debye1947}:
\begin{equation}
F(q)=\frac{2}{u^{2}}\left(e^{-u}+u-1\right),\qquad u=(qR_g)^{2},\qquad R_g^{2}=\frac{L\,l_p}{3},
\end{equation}
where $L$ and $l_p$ are the polymer's contour and persistence lengths, respectively. Therefore, one may distinguish between two polydispersity scenarios: various polymer length distribution as in the case of synthetic polymer production, or alternative polymer conformations of polymers of identical length. We followed the latter class, representing conformations of intrinsically disordered proteins \cite{svergun2003,koren_not_2026,aodeh_neurofilament_2026,kravikass_isolated_2024}. 

\begin{figure}
\includegraphics[width=0.8\textwidth]{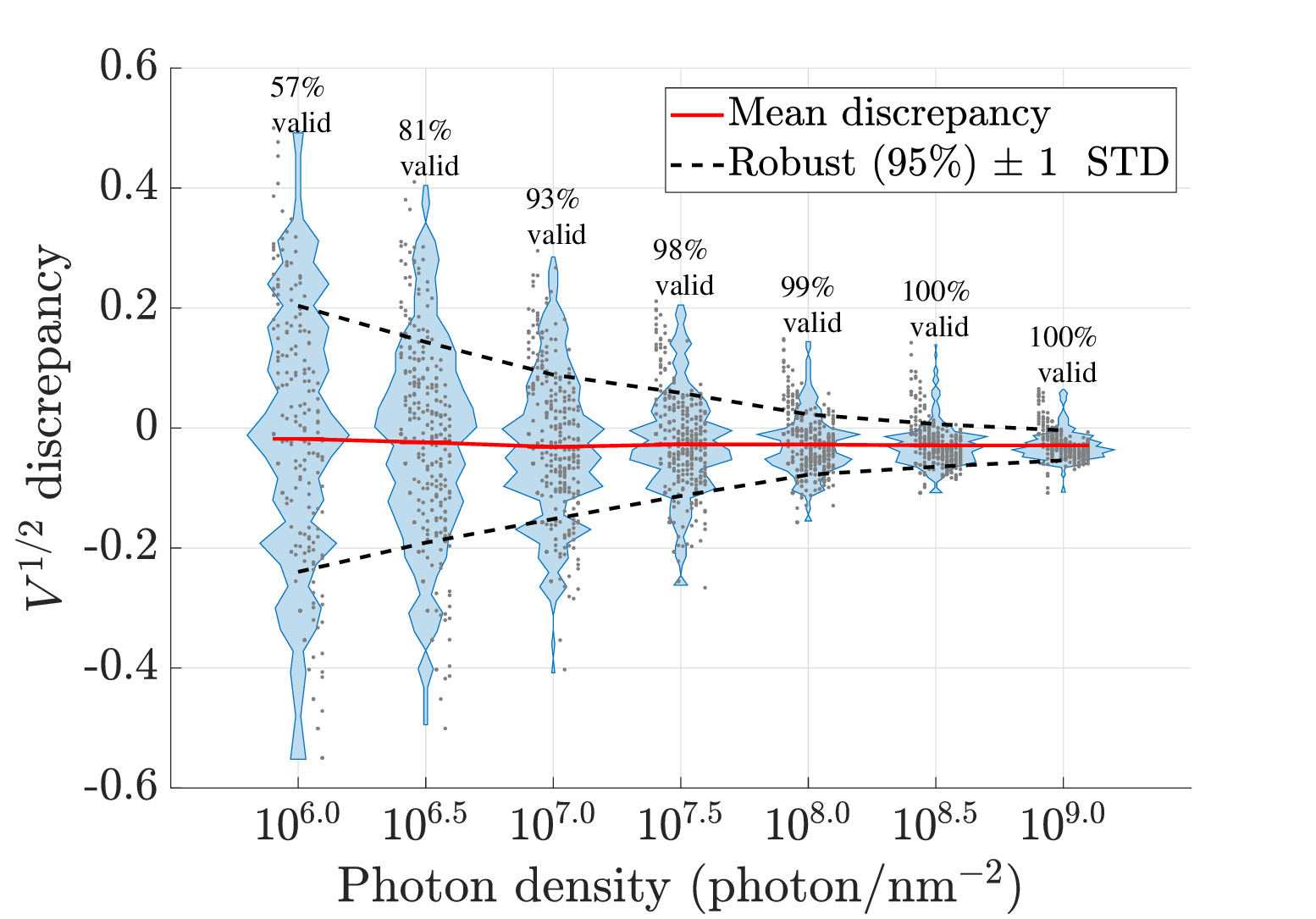}
\caption{The discrepancy statistics of simulations of ensembles of Gaussian chain form factor, at different collected photon densities. The ensemble's radius of gyration distribution is normal with a mean of $R_0=3\,\mathrm{nm}$ and relative standard deviations in the range of 0-0.55. The $\mathcal{Y}_G$ obtained from each simulation were compared with the theoretical formula. In this violin plot the patches represent the histogram of the discrepancy of the derived relative standard deviation ($V^{1/2}$) from the actual value for each signal strength (the scattering weighting for this case is trivial). 5\% outliers were omitted from the analysis. The mean bias is represented by a solid red line, while the discrepancy standard deviation margin is indicated by the dashed line. The percentage of valid points (where the realization yielded a finite estimate within the admissible range of the analytical inversion) is indicated on each histogram.}
\label{fig:V-discrep}
\end{figure}  

\begin{figure}
\includegraphics[width=0.8\textwidth]{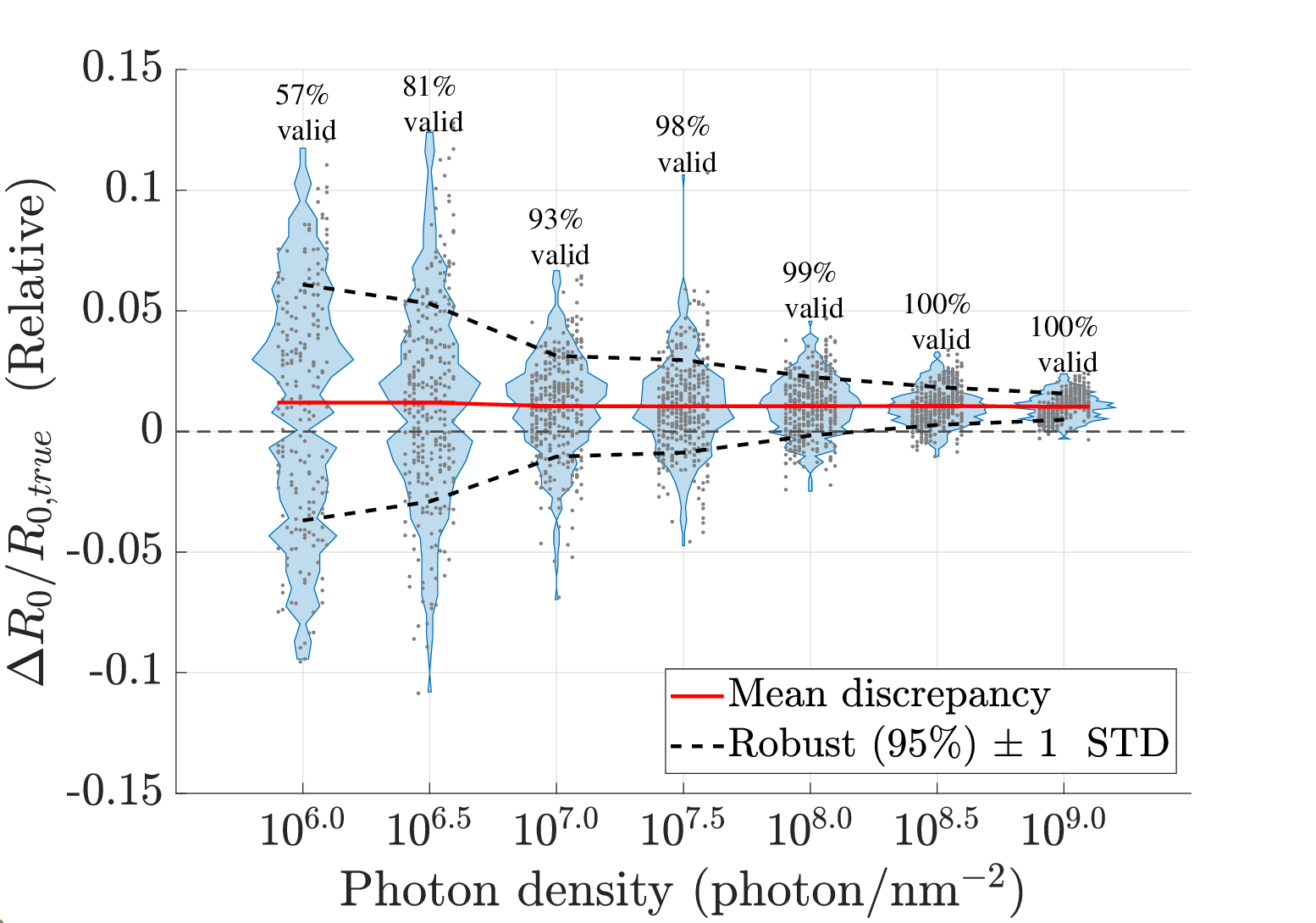}
\caption{Relative discrepancy of the recovered scattering-weighted mean radius
$R_0=\tilde{R}_0/\sqrt{1+V_{est}}$ against its true ensemble value, as a function of the accumulated scattered-photon density $N_{q^{2}}$. Extracting $R_0$ requires the variance, so this discrepancy inherits the uncertainty of $V_{est}$; it is the quantity of physical interest, as opposed to the apparent Guinier radius $\tilde{R}_0$, which an ordinary Guinier analysis already returns. The simulation conditions are detailed in Tab.\,\ref{tab:beam_simul_param}, and demonstrate Gaussian-chain form factor with normal $R_g$ distribution (this figure is based on the same dataset used in Fig.\,\ref{fig:V-discrep}). The radius is recovered to an RMS relative accuracy of $2.0\%$ at $N_{q^2}=10^{7}\, \mathrm{photons\,nm^2}$, approaching a systematic floor of ${+}1.0\%$ inherited from the residual bias of $V_{est}$. Grey points are individual replicates, the red line the 5\%-trimmed mean and the dashed lines $\pm1$ trimmed standard deviation; the percentage of runs returning a finite estimate ('valid') is annotated per column.}
\label{fig:Rg-discrep}
\end{figure}

For such a model, the scattering strength is independent of $R_g$ so the number and the scattering-weighted size distributions coincide. As expected, for different numbers of photons (and hence noise level) we see an increase in performance as the noise level decreases. We estimated the variance \(V_{est}\) based on the $\mathcal{Y}_G$ observable, by comparing to the analytical function (Fig.\,\ref{fig:V-discrep}). The results show that when the peak collected forward scattered photon density in the relevant $q$-space exceeds $10^7\,\mathrm{photons\,nm^{2}}$, over 90\% of the measurements show valid variances (the observable is within the analytical range). The estimated distribution spread (with 5\% outlier omission) is slightly biased ($\left<\Delta V^{1/2}\right>\approx0.03$), with a discrepancy standard deviation of about 0.1 in $V^{1/2}$. The background subtracted photons needed for this performance are collected within a fraction of a typical exposure (one second of measurement) in the Diamond light source BioSAXS B21 beamline with a regular sample, making TENOR-SAXS a feasible method (Appendix \ref{sensitivityAnalisys}). At the same photon density the scattering-weighted average Guinier radius ($R_0=\tilde{R}_0/\sqrt{1+V_{est}}$) is recovered with ${\sim}2\%$ RMS relative precision (Fig.\,\ref{fig:Rg-discrep}).

\section{Summary and Conclusions}
We presented here a \textbf{T}echnique for \textbf{En}semble \textbf{O}bservation by \textbf{R}esolution variation in \textbf{SAXS} (TENOR-SAXS). It provides a practical means to quantify ensemble variance directly from two-dimensional scattering images while removing explicit dependence on instrumental resolution functions. By exploiting digital smearing, the method retrieves distributional information in polydisperse samples that is not accessible through conventional Guinier analysis. In contrast to ensemble-modeling strategies that fit large structural pools to one-dimensional profiles~\cite{bernado2007,tria2015,antonov2016}, or density-reconstruction approaches that recover three-dimensional real-space maps~\cite{grant2018}, TENOR-SAXS uses controlled resolution variation to construct observables which are only weakly sensitive to the native PSF. The technique thus enables ensemble characterization without imposing heavy structural priors, complementing both model-based and model-free SAXS methodologies~\cite{hura2009,rambo2013,burger2016}.
TENOR-SAXS offers two principal advantages: practicality and robustness. Digital smearing eliminates the need for repeated measurements under modified beamline conditions, thereby reducing beamtime requirements, alignment and normalization errors. Moreover, the method requires no assumption about the shape of the size distribution and no knowledge of the instrumental resolution function. Nonetheless, it does require the knowledge of the monodisperse form factor, or at least its Guinier curvature $\phi''$. 

Our results indicate that systematic deviations remain negligible under realistic conditions, with ensemble relative standard deviation recovered within $\Delta V^{1/2} < 0.1$ when scattered-photon densities exceed $N_{q^{2}}\sim10^{7}$~photons\,nm$^{2}$. Such levels are readily achievable at modern synchrotron facilities and accessible for advanced laboratory SAXS instruments. Because the analysis operates directly on raw 2D intensity maps, the method may also be applied retrospectively to archived datasets for which the raw 2D detector images have been retained, broadening the interpretative value of existing data. We note that community repositories such as SASBDB \cite{kikhney2020sasbdb} archive azimuthally averaged profiles rather than detector frames, so retrospective application depends on raw 2D images retained by the facility or the depositor.

The approach has some limitations. Its current formulation applies to the Guinier regime, requires modest anisotropic differences in smearing, and depends on the curvature term in the Guinier expansion ($\phi''$). Noise sensitivity increases at low photon counts, sometimes preventing the variance from being extracted from the measurement. However, the noise dependency is predictable and allows experiments to be designed accordingly. Buffer subtraction, typically performed after azimuthal averaging, must be carried out consistently at the 2D level to avoid introducing anisotropic artifacts to the log-ratio observables. Finally, for sufficiently broad distributions, higher distribution moments, particularly skewness, begin to influence the observables. For the physically admissible distributions examined here, however, the attainable skewness is limited, and its effect does not substantially restrict the working range of the method.

Several avenues for further development are apparent. Alternative observables may reduce dependence on $\phi''$ and enhance noise suppression. Extension to higher $q$-ranges and more complex form factors would broaden applicability. Systematic mapping of dependencies on point-spread function size, detector resolution, and fitting strategies may provide practical guidelines for experiment design. Incorporating independent structural information, such as externally determined $\phi''$ \cite{bennun2010xplus,ginsburg2019dplus} or orthogonal fitted parameters, may further reduce uncertainty. Further analysis is needed for practical conformations where the form factor curvature is related to the radius of gyration. The code package is available for researchers to use and implement on experimental data \cite{TENORSAXS_GitHub}.

In summary, TENOR-SAXS introduces a conceptually simple, PSF-robust framework for ensemble characterization in SAXS. By combining practical implementation, robustness to systematic effects, and complementarity with existing approaches, it has the potential to substantially expand the interpretative power of small-angle scattering in both biological and materials sciences.
\\

\section{Code and data availability}
Both MATLAB and Python codes reproducing the simulations and analysis are available in the TENOR-SAXS GitHub repository which includes an interactive online implementation and is also available as a Google Colab notebook \cite{TENORSAXS_GitHub}. 
\appendix
\\
\\

\section{Direct derivation of the second order log-intensity}
\label{app:direct-logI}

We start from the definition of the measured intensity as a convolution of the intrinsic form factor with the instrumental PSF:
\begin{equation}
\begin{split}
I(\mathbf q) \;=\; R (q) \circledast F(q) \;=\; \int R(-\mathbf p)\,F(\mathbf q+\mathbf p)\,d^2p,\\
\int R(\mathbf p)\,d^2p=1.
\end{split}
\label{eq:conv-direct}
\end{equation}
Assume \(R\) is a centered, separable anisotropic Gaussian,
\begin{equation}
\begin{aligned}
R(\mathbf q) \;=\; \frac{1}{2\pi\sigma_x\sigma_y}\,
\exp\!\Big(-\frac{q_x^2}{2\sigma_x^2}-\frac{q_y^2}{2\sigma_y^2}\Big),
\qquad\\
\langle q_x\rangle=\langle q_y\rangle=0,\;\;
\langle q_x^2\rangle=\sigma_x^2,\;\;\langle q_y^2\rangle=\sigma_y^2,\;\;\langle q_xq_y\rangle=0.
\label{eq:R-gauss-direct}
\end{aligned}
\end{equation}

Expand \(F(\mathbf q+\mathbf p)\) to second order:
\begin{equation}
\begin{split}
F(\mathbf q+\mathbf p) \;\approx\;& F(\mathbf q)
+ \sum_i{p_i\,\partial_{q_i}F(\mathbf q)}\\
&+ \frac{1}{2}\sum_{i,j}{p_ip_j\,\partial_{q_i}\partial_{q_j}F(\mathbf q)},
\qquad i,j\in\{x,y\}.
\end{split}
\label{eq:taylor-F}
\end{equation}
Averaging against the even kernel \(R\) eliminates the linear term. Using the second moments in Eq.~\eqref{eq:R-gauss-direct} gives
\begin{equation}
I(\mathbf q) \;\approx\; F(\mathbf q) \;+\; \frac{1}{2}\Big[\sigma_x^2\,\partial_{q_x}^2+\sigma_y^2\,\partial_{q_y}^2\Big]F(\mathbf q).
\label{eq:I-second-order}
\end{equation}
This is the standard small-PSF result. More generally, for any centered kernel with covariance matrix \(C\), one has \(I \approx F + \tfrac{1}{2}C_{ij}\partial_{q_i}\partial_{q_j}F\).

Next, we write Eq.~\eqref{eq:I-second-order} as \(I=F+\delta\) with
\begin{equation}
\frac{\delta}{F} \;=\; \frac{1}{2}\left[\sigma_x^2\,\frac{\partial_{q_x}^2F}{F}+\sigma_y^2\,\frac{\partial_{q_y}^2F}{F}\right].
\label{eq:delta-over-F}
\end{equation}
Following, we define the dimensionless smallness measure as:
\begin{equation}
\begin{split}
\varepsilon(\mathbf q) \;\equiv\; \max\Big\{\,&
\big|\sigma_x^2\,\partial_{q_x}^2\ln F(\mathbf q)\big|,\;
\big|\sigma_y^2\,\partial_{q_y}^2\ln F(\mathbf q)\big|,\; \\
&\big|\sigma_x^2\big(\partial_{q_x}\ln F(\mathbf q)\big)^2\big|,\;
\big|\sigma_y^2\big(\partial_{q_y}\ln F(\mathbf q)\big)^2\big|
\,\Big\} \;\ll\; 1,
\end{split}
\end{equation}
so that \(\delta/F = \mathcal O\!\big(\varepsilon(\mathbf q)\big)\) is dimensionless. Since $\ln(1+x)=x+\mathcal {O}(x^2)$, the expanded log scattered intensity becomes: 
\begin{equation}
\ln I(\mathbf q) \;=\; \ln F(\mathbf q) + \ln\!\Big(1+\frac{\delta}{F}\Big)
\;\approx\; \ln F(\mathbf q) + \frac{\delta}{F}
\;+\; \mathcal O\!\big(\varepsilon(\mathbf q)^2\big).
\label{eq:logI-expansion}
\end{equation}
Using the identity
\begin{equation}
\frac{1}{F}\,\partial_{q_i}^2F \;=\; \partial_{q_i}^2\ln F \;+\; \big(\partial_{q_i}\ln F\big)^2,
\qquad i\in\{x,y\},
\end{equation}
and substituting into Eq.\,\eqref{eq:logI-expansion} yields the second order log-intensity formula in a fully dimensionless form:

\begin{equation}
\begin{aligned}
\ln I(\mathbf q)\;\approx\;&\;\ln F(q) + \tfrac{1}{2}\big(\sigma_x^2\,\partial_{q_x}^2+\sigma_y^2\,\partial_{q_y}^2\big)\ln F(q) \\[6pt]
&+ \tfrac{1}{2}\Big[\sigma_x^2\big(\partial_{q_x}\ln F(q)\big)^2
 + \sigma_y^2\big(\partial_{q_y}\ln F(q)\big)^2\Big] \\ &+ \mathcal O\!\big(\varepsilon(\mathbf q)^2\big)
\end{aligned}
\label{eq:logI-operator-appendix-updated}
\end{equation}

The neglected terms are of second order in the dimensionless combinations \(\sigma_i^2\,\partial_{q_i}^2\ln F\) and \(\sigma_i^2\,(\partial_{q_i}\ln F)^2\). This is met when the angular smearing is smaller than the form factor's second log-$q$-derivative which is typically comparable to $R_g$. That is, the PSF requirement for this approximation is $2\sigma_i^2 R_0 ^2/3 \ll 1$ at the Guinier regime.

For an isotropic form factor \(F(\mathbf q)=F(| \mathbf q|)\), define the first and second radial derivatives
\[
f(q)\equiv \partial_q\ln F(q),\qquad f'(q)\equiv \partial_q^2\ln F(q).
\]

Projecting the Cartesian derivatives onto the radial direction then yields for $i\in\{x,y\}$,
\begin{align} 
\partial_{q_i}\ln F &= f(q)\frac{q_i}{q}, \\[3pt] \partial^2_{q_i}\ln F &= f'(q)\frac{q_i^2}{q^2} +\frac{f(q)}{q} \left( 1 -\frac{q_i^2}{q^2} \right). 
\end{align}
Next, we define the azimuth $\chi$ by $(\frac{q_x}{q}, \frac{q_y}{q})=(\cos\chi,\sin\chi)$ and the effective radial smearing at each point by: 
\begin{align}
\Sigma_{\mathrm{eff},k}^2(\chi)
&\equiv\sigma_{x,k}^2\cos^2\chi+\sigma_{y,k}^2\sin^2\chi\ ,
\label{eq:sigma_1}
\end{align} where $k$ enumerates different PSF shapes.
Following, we can now transform Eq.\,\eqref{eq:logI-operator-appendix-updated} to obtain the second-order result in polar form:
\begin{equation}
\begin{split}
\ln I_k(q,\chi)\approx \ln F(q)+\frac{1}{2}\Big[
&\Big(f'(q)+f(q)^2\Big)\,\Sigma_{\mathrm{eff},k}^2(\chi)\\
&+\frac{f(q)}{q}\Big(\sigma_{x,k}^2+\sigma_{y,k}^2-\Sigma_{\mathrm{eff},k}^2(\chi)\Big)
\Big].
\end{split}
\label{eq:PSF-cart-core}
\end{equation}

Next, from geometrical considerations ($\cos^2\chi=(1+\cos 2\chi)/2$ and $\sin^2\chi=(1-\cos 2\chi)/2$), we can rewrite Eq.\,\eqref{eq:sigma_1}:
\begin{equation}
\begin{split}
\Sigma_{\mathrm{eff},k}^2(\chi)&=\bar\sigma_k^2+\frac{\Delta\sigma_k^2}{2}\cos 2\chi,\\
\sigma_{x,k}^2+\sigma_{y,k}^2-\Sigma_{\mathrm{eff},k}^2(\chi)&=\bar\sigma_k^2-\frac{\Delta\sigma_k^2}{2}\cos 2\chi,
\end{split}
\end{equation}
where
\begin{equation}
\bar\sigma_k^2=\frac{\sigma_{x,k}^2+\sigma_{y,k}^2}{2},\qquad \Delta\sigma_k^2=\sigma_{x,k}^2-\sigma_{y,k}^2.
\end{equation}
Next, we define the PSF scalars:
\begin{equation}
\Sigma_{0,k}\equiv \bar\sigma_k^2,\qquad \Delta_{0,k}\equiv \frac{\Delta\sigma_k^2}{2},
\label{eq:Sigma0-Delta0}
\end{equation}
and the two radial response functions:
\begin{equation}
G(q)\equiv \frac{f(q)}{q}+f'(q)+f(q)^2,\qquad
M(q)\equiv -\frac{f(q)}{q}+f'(q)+f(q)^2.
\label{eq:G-M-defs}
\end{equation}
Then Eq.\,\eqref{eq:PSF-cart-core} becomes the pixel-wise model
\begin{equation}
\begin{split}
\ln I_k(q,\chi)&\approx \ln F(q)+\mathcal{G}_k(q,\chi),\quad\text{where}\\
\mathcal{G}_k(q,\chi)&=\frac{1}{2}\,\Sigma_{0,k}\,G(q)+\frac{1}{2}\,\Delta_{0,k}\,M(q)\cos 2\chi
\end{split}
\label{eq:logIk}
\end{equation}
which is the form used for the 2D parameter extraction. Specifically, for two measurements with different PSFs,
subtracting Eq.\,\eqref{eq:logIk} eliminates $\ln F$ and yields:
\begin{equation}
\begin{split}
\big[\ln I\big]_1-\big[\ln I\big]_2
&=\frac{1}{2}\Big(\Sigma_{0,1}-\Sigma_{0,2}\Big)\,G(q)\\
&\quad+\frac{1}{2}\Big(\Delta_{0,1}-\Delta_{0,2}\Big)\,M(q)\cos 2\chi\equiv\mathcal{G}(q,\chi),
\end{split}
\label{eq:G-cart}
\end{equation}
which separates the angle independent and $\cos 2\chi$ components and depends only on the differences of the PSF scalars and the common radial functions $G$ and $M$.

\section{Cumulant expansion of the ensemble-averaged intensity}
\label{app:cumulant}

\subsection{Notation and dimensionless variables}
We consider isotropic single-particle scattering, so the intrinsic (monodisperse) form factor depends on the radial scattering vector only via $q^2$. For an ensemble with a distribution of radii of gyration $R_g$, we define the random variable $X \equiv R_g^2$, and the dimensionless argument $u \equiv q^2 X$. \\

Let $\phi(u)$ denote the \emph{monodisperse} log-form factor,
\[
\ln F(q\,|\,X) \equiv \phi(u),
\]
assumed analytic at $u=0$. Its Taylor series coefficients at the origin are
\begin{equation}
\begin{split}
\phi(u)&= \kappa\,u + \tfrac{1}{2}\phi''\,u^2 + \tfrac{1}{6}\phi^{(3)}\,u^3 + \mathcal{O}(u^4),
\\
\kappa&=\partial\phi(0),\;\; \phi''=\partial^2\phi(0),\;\; \phi^{(3)}=\partial^3\phi(0),
\end{split}
\label{eq:phi_series_app}
\end{equation}
and relative to the intensity at $q=0$, the ensemble-averaged measured intensity is: 
\begin{equation}
\begin{split}
I(q)=\mathbb{E}_X\!\Big[\exp\!\big(\phi(q^2 X)\big)\Big] = \mathbb{E}_X\!\big[e^{\,\phi(u)}\big], 
\\
\ln I(q)=\ln \mathbb{E}_X\!\big[e^{\,\phi(u)}\big].
\label{eq:def_lnI_app}
\end{split}
\end{equation}
\subsection{Cumulant-generating function}
We introduce the random variable $Z \equiv \phi(u)$ and its cumulant-generating function (CGF):
\begin{equation}
K_Z(s)=\ln \mathbb{E}[e^{s Z}].
\end{equation}
Following, $\ln I(q)=K_Z(1)$ and therefore we can expand the logarithm such that: 
\begin{equation}
\ln I(q)=\sum_{n\ge 1}\frac{\kappa_n(Z)}{n!}
=\underbrace{\mathbb{E}[Z]}_{\kappa_1(Z)}+\frac{1}{2}\underbrace{\mathrm{Var}(Z)}_{\kappa_2(Z)}
+\frac{1}{6}\kappa_3(Z)+\mathcal{O}(u^4).
\label{eq:cum_series_app}
\end{equation}
Using Eq.\,\eqref{eq:phi_series_app} up to $\mathcal{O}(u^3)$ we can rewrite our random variable as: 
\begin{equation}
Z = A\,u + B\,u^2 + C\,u^3 + \mathcal{O}(u^4),
\qquad
A=\kappa,\quad B=\tfrac{1}{2}\phi'',\quad C=\tfrac{1}{6}\phi^{(3)}.
\label{eq:Z_ABC_app}
\end{equation}

According to cumulant-expansion theorem, we can express the cumulants using Eq.\,\eqref{eq:Z_ABC_app}. The first cumulant is therefore: 
\begin{equation}
\kappa_1(Z)=\mathbb{E}[Z]
= A\,\mathbb{E}[u] + B\,\mathbb{E}[u^2] + C\,\mathbb{E}[u^3] + \mathcal{O}(u^4).
\label{eq:k1_app}
\end{equation}
Evaluating the expansion through $\mathcal{O}(u^3)$ requires $Z$ through cubic order while the second-cumulant requires only the linear and quadratic terms in $Z$:
\begin{equation}
\kappa_2(Z)=\mathrm{Var}(Z)=A^2\,\mathrm{Var}(u)+2AB\,\mathrm{Cov}(u,u^2) + \mathcal{O}(u^4).
\label{eq:varZ_app}
\end{equation}
And last, to a leading term of $\mathcal{O}(u^3)$, the third cumulant exists even when $C= 0$ and can be expressed as:
\begin{equation}
\begin{split}
\kappa_3(Z)&=\mathbb{E}\!\big[(Z-\mathbb{E}[Z])^3\big]\\
&=\mathbb{E}\big[(A u - A\,\mathbb{E}[u])^3\big] + \mathcal{O}(u^4)
= A^3\,\kappa_3(u) + \mathcal{O}(u^4).
\end{split}
\end{equation}
For a given $q$, the random variable is $X$, and therefore $\kappa_3(u)=q^6\,\kappa_3(X)$. We argue (see App.\,\ref{sensitivityAnalisys}) that in realistic polydisperse size distributions the scattering-weighted skewness is limited, and $\kappa_3(X)=\mathcal{O}(R_0^6V^{3/2})$, hence
\begin{equation}
\kappa_3(Z) = \mathcal{O}\!\big(u^3\,V^{3/2}\big).
\label{eq:k3_drop_app}
\end{equation}
\\
Next, we evaluate the leading raw moments in terms of $X$ for a given $q$:
\begin{equation}
\begin{aligned}
\mathbb{E}[u]=q^2\,\mu_1,\quad \mathbb{E}[u^2]=q^4\,\mu_2,\quad \mathbb{E}[u^3]=q^6\,\mu_3, 
\qquad \\
\mathrm{Var}(u)=q^4\,\mathrm{Var}(X),\quad \mathrm{Cov}(u,u^2)=q^6\,\mathrm{Cov}(X,X^2),
\label{eq:u_moments_app}
\end{aligned}
\end{equation}
where $\mu_k\equiv \mathbb{E}[X^k]$. Thus, even if $\phi^{(3)}=0$ (i.e., $C=0$), the $\mathcal{O}(u^3)$ contribution survives. However, for small values of $V$, the component originating from distribution skewness can be neglected. For a narrow ensemble of $R_g$ with mean $R_0=\mathbb{E}[R_g]$ and dimensionless variance $V=\mathrm{Var}(R_g)/R_0^2\ll 1$, the required combinations to first order in $V$ are
\begin{equation}\begin{split}
\mu_1=R_0^2(1+V),\\ \mu_2=R_0^4(1+6V)+\mathcal{O}(V^{3/2}),
\\ \mu_3=R_0^6(1+15V)+\mathcal{O}(V^{3/2}),
\end{split}\label{eq:X_moments_app}
\end{equation}
and 
\begin{equation}
\mathrm{Var}(X)=4R_0^4V+\mathcal{O}(V^{3/2}),\qquad 
\mathrm{Cov}(X,X^2)=8R_0^6V+\mathcal{O}(V^{3/2}).
\label{eq:cov_app}
\end{equation}

\subsection{Result for $\ln I(q)$ through $\mathcal{O}(u^3)$}
Combining \eqref{eq:cum_series_app}-\eqref{eq:cov_app} and keeping terms through $\mathcal{O}(u^3)$ and linear in $V$,

\begin{align}
\ln I(q)
&= A\,\mathbb{E}[u] + B\,\mathbb{E}[u^2] + \frac{1}{2}A^2\,\mathrm{Var}(u) + A B\,\mathrm{Cov}(u,u^2) 
+ \mathcal{O}(u^4) \nonumber\\[2pt]
&= \kappa\,q^2\,\mu_1
\;+\; \tfrac{1}{2}\phi''\,q^4\,\mu_2
\;+\; \tfrac{1}{2}\kappa^2\,q^4\,\mathrm{Var}(X) \nonumber\\[2pt]
&\quad \;+\; \tfrac{1}{2}\kappa\phi''\,q^6\,\mathrm{Cov}(X,X^2)
\;+\; \mathcal{O}(u^4).
\label{eq:lnI_u_app}
\end{align}
Using \eqref{eq:X_moments_app}-\eqref{eq:cov_app} and defining $Q\equiv q^2 R_0^2$,

\begin{equation} \begin{aligned} \ln I(q) ={}& \kappa(1+V)Q +\left[ \tfrac12\phi''(1+6V) +2\kappa^2V \right]Q^2 \\ &+\left[  4\kappa\phi''V + \tfrac16\phi^{(3)}(1+15V) \right]Q^3 +\mathcal{O}\!\left(V^{3/2}Q^2\right) \\ &+\mathcal{O}\!\left(V^{3/2}Q^3\right) +\mathcal{O}\!\left(Q^4\right), \end{aligned} \label{eq:lnI_dimless_app} \end{equation}

which limits the analysis to sub-Guinier regime ($qR_0<1$). We can further rewrite Eq.\,\eqref{eq:lnI_dimless_app} as a polynomial in $Q$: 
\begin{equation}
\begin{split}
\ln I(q) \;=\;& a\,Q \;+\; b\,Q^2 \;+\; c\,Q^3\\
&+\;\mathcal{O}\!\left(V^{3/2}Q^2\right)\; +\;\mathcal{O}\!\left(V^{3/2}Q^3\right) +\; \mathcal{O}(Q^4),
\end{split}
\label{eq:abc_Q_app}
\end{equation}
where, to first order in $V$:

\begin{align}
a &= \kappa\,(1+V), \\
b &= \tfrac{1}{2}\phi''\,(1+6V) + 2\kappa^2\,V, \\
c &= 4\kappa\phi'' \, V + \tfrac16\phi^{(3)}(1+15V). 
\end{align}

\subsection{From cumulant expansion to $G$ and $M$ functions}

We are now in position to express the polynomial functions $G(q)$ and $M(q)$ in Eq.\,\eqref{eq:GM-poly} using the cumulant expansion in Eq.\,\eqref{eq:abc_Q_app}. 
In the main text, we identified the polynomial function (Eq.\,\eqref{eq:GM_defs}) as 
\(
G(q)=f'(q)+\tfrac{f(q)}{q}+f(q)^2
\)
and 
\(
M(q)=f'(q)-\tfrac{f(q)}{q}+f(q)^2
\),
where $f(q)=\partial_{q}\ln I(q)$. Using Eq.\,\eqref{eq:abc_Q_app} and $Q=q^2 R_0^2$, the expansions in powers of $q^2$ are
\begin{align}
G(q) &= 4a\,R_0^2 \;+\; (16b+4a^2)\,R_0^4\,q^2 \;+\; (36c+16ab)\,R_0^6\,q^4 \notag\\
&\quad \;+\; (24ac+16b^2)\,R_0^8\,q^6 \;+\; \mathcal{O}(q^8), \label{eq:G_from_abc_app}\\
M(q) &= 0 \;+\; (8b+4a^2)\,R_0^4\,q^2 \;+\; (24c+16ab)\,R_0^6\,q^4 \notag\\
&\quad \;+\; (24ac+16b^2)\,R_0^8\,q^6 \;+\; \mathcal{O}(q^8). \label{eq:M_from_abc_app}
\end{align}
Equations \eqref{eq:lnI_dimless_app}-\eqref{eq:M_from_abc_app} provide the dimensionless, order-by-order link between the ensemble cumulants (via $a,b,c$) and the measured 2D coefficients.

We keep only linear $V$ terms (no skewness), and remove terms including $\phi^{(3)}$ (for the sake of simplicity and not wanting to rely too much on knowledge of the exact form factor). We can now use the cumulant expansion to identify the relation between $a,b,c$ and the polynomial coefficients for $G(q)$ and $M(q)$ shown in Tab.\ \ref{tab:GM-coef} where we show expressions truncated at $\mathcal{O}(V)$ and $\mathcal{O}(q^6)$. The $\phi^{3}$ omission affects only the coefficients of orders higher than $q^4$. In general the results presented in Tab.\,\ref{tab:GM-coef} are a good first approximation to many cases, but appropriate cumulant expansion should be practiced for the relevant form factor under investigation. In practice, given a relevant form factor, one can simulate the ideal scattering for various $V$ to find the relevant coefficients $G(q,V)$ and $M(q,V)$. This can be used to produce a calibration curve from the observables (Eqs. \eqref{eq:observables}-\eqref{eq:R_m-observe}) to $V$, especially the observables involving the higher order $q$ terms. 
\begin{figure}
\includegraphics[width=1\textwidth]{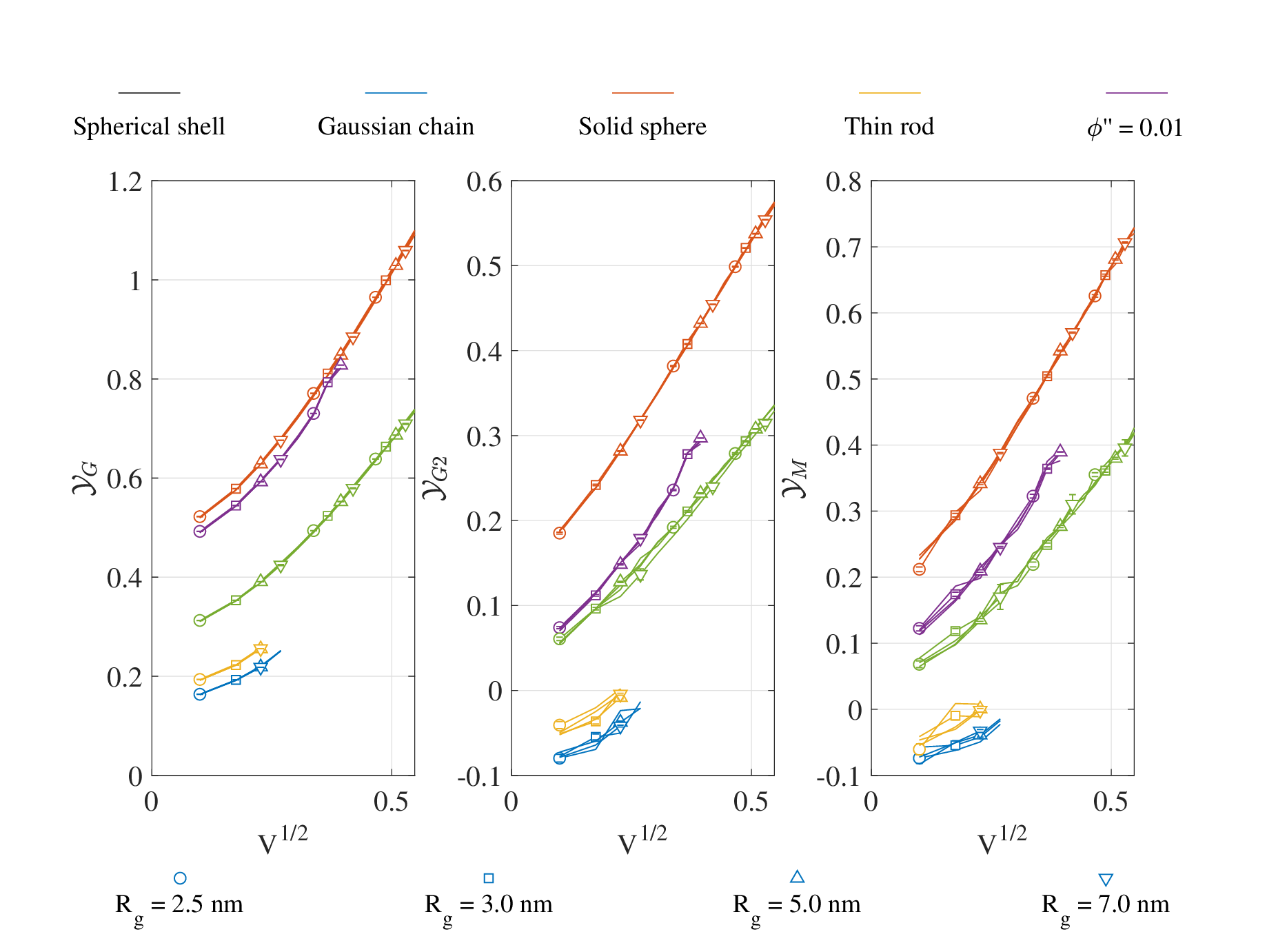}
\caption{Sensitivity of the $\mathcal{Y}$ observables to the weighted average radius of gyration.}
\label{fig:sensitiVT}
\end{figure}
\begin{figure}
\includegraphics[width=1\textwidth]{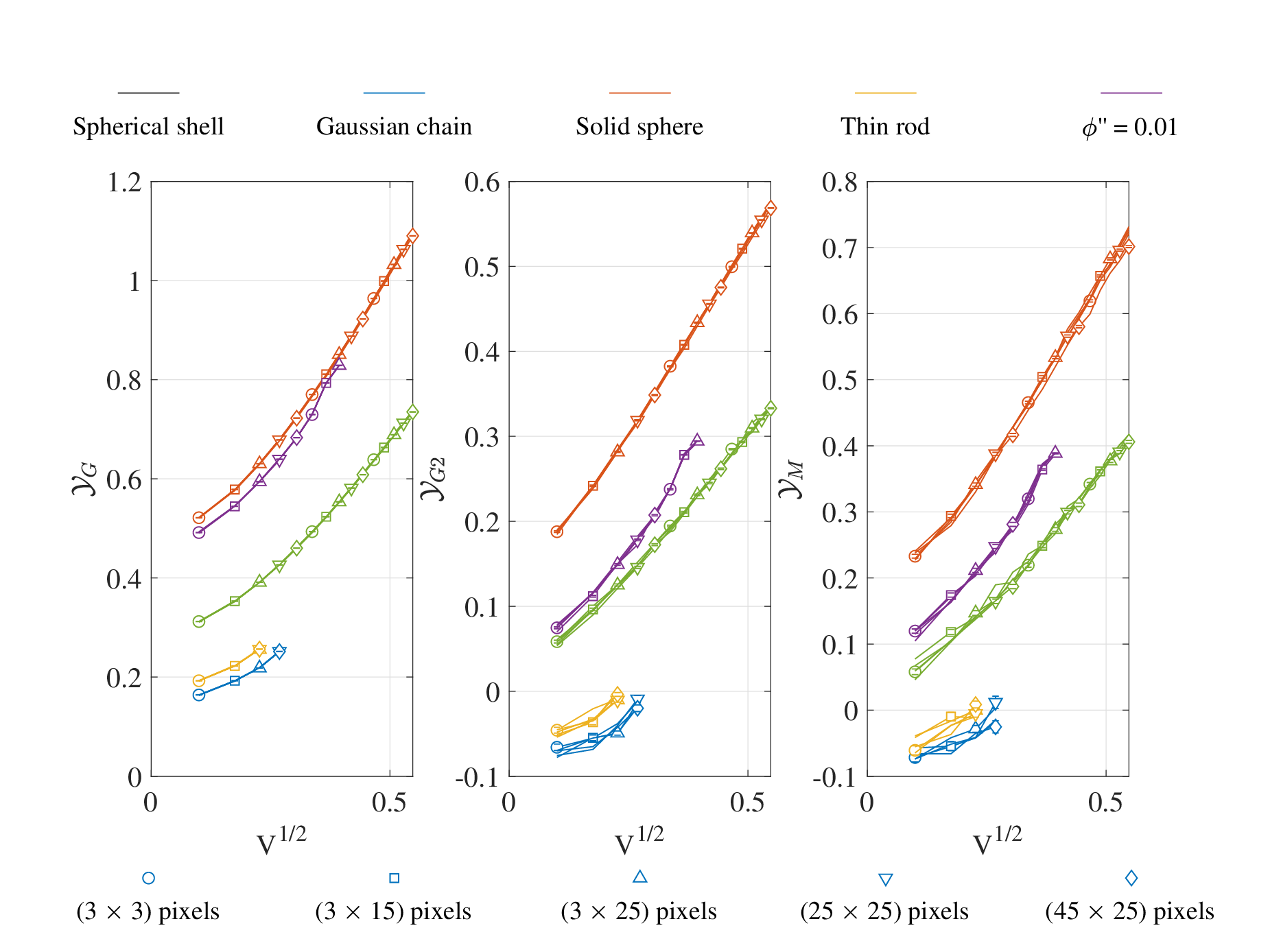}
\caption{Sensitivity to different instrumental PSF sizes (expressed in its footprint's number of detector pixels) in the extraction of the observables.}
\label{fig:sensitiVT3}
\end{figure}

\section{Degree of independence on ensemble and instrumental parameters}
\label{sensitivityAnalisys}
\begin{figure}
\includegraphics[width=1\textwidth]{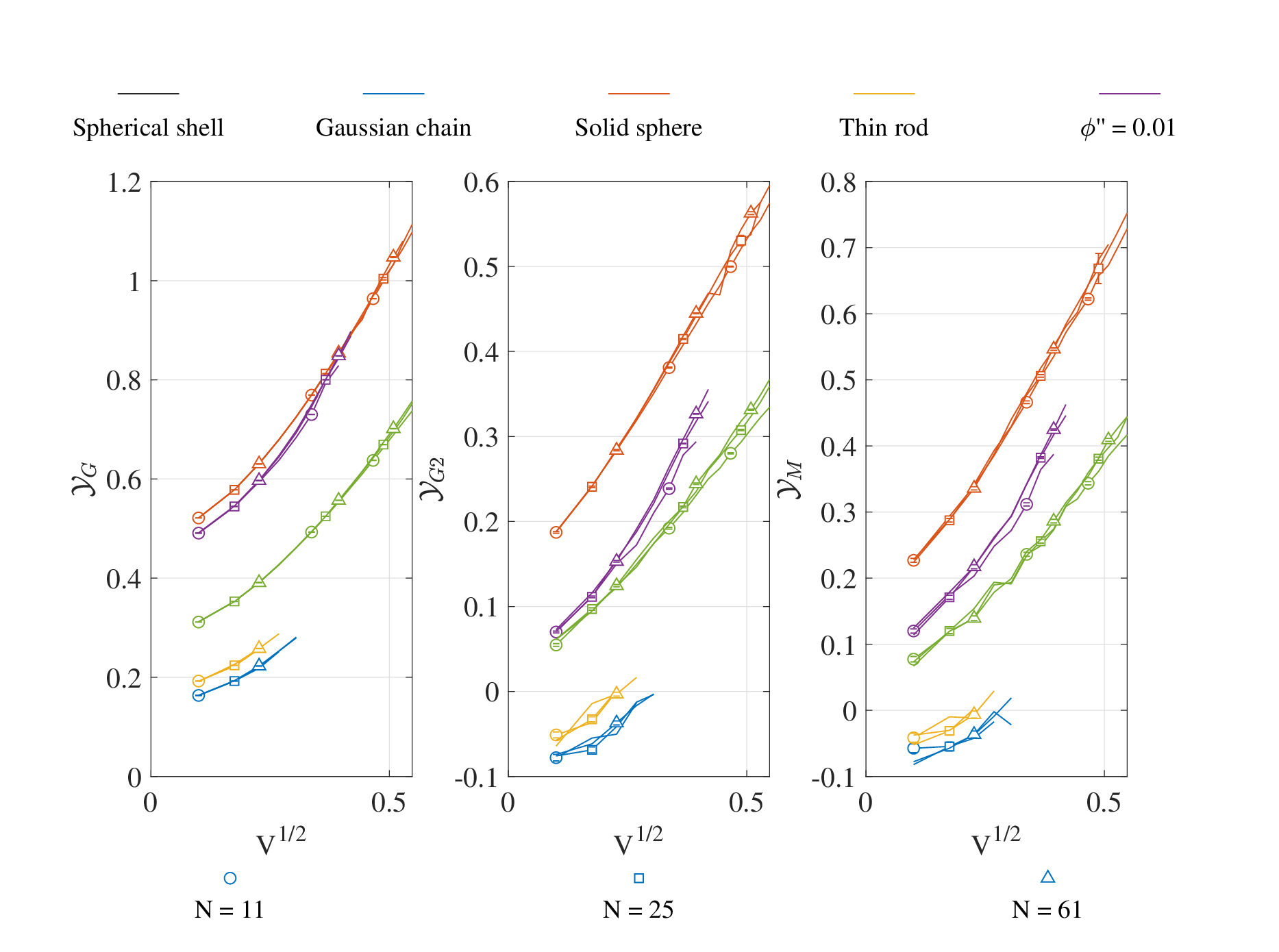}
\caption{Sensitivity to the simulated heterogeneity, the number of different radii used.}
\label{fig:sensitiVT1}
\end{figure}

\begin{figure}
\includegraphics[width=1\textwidth]{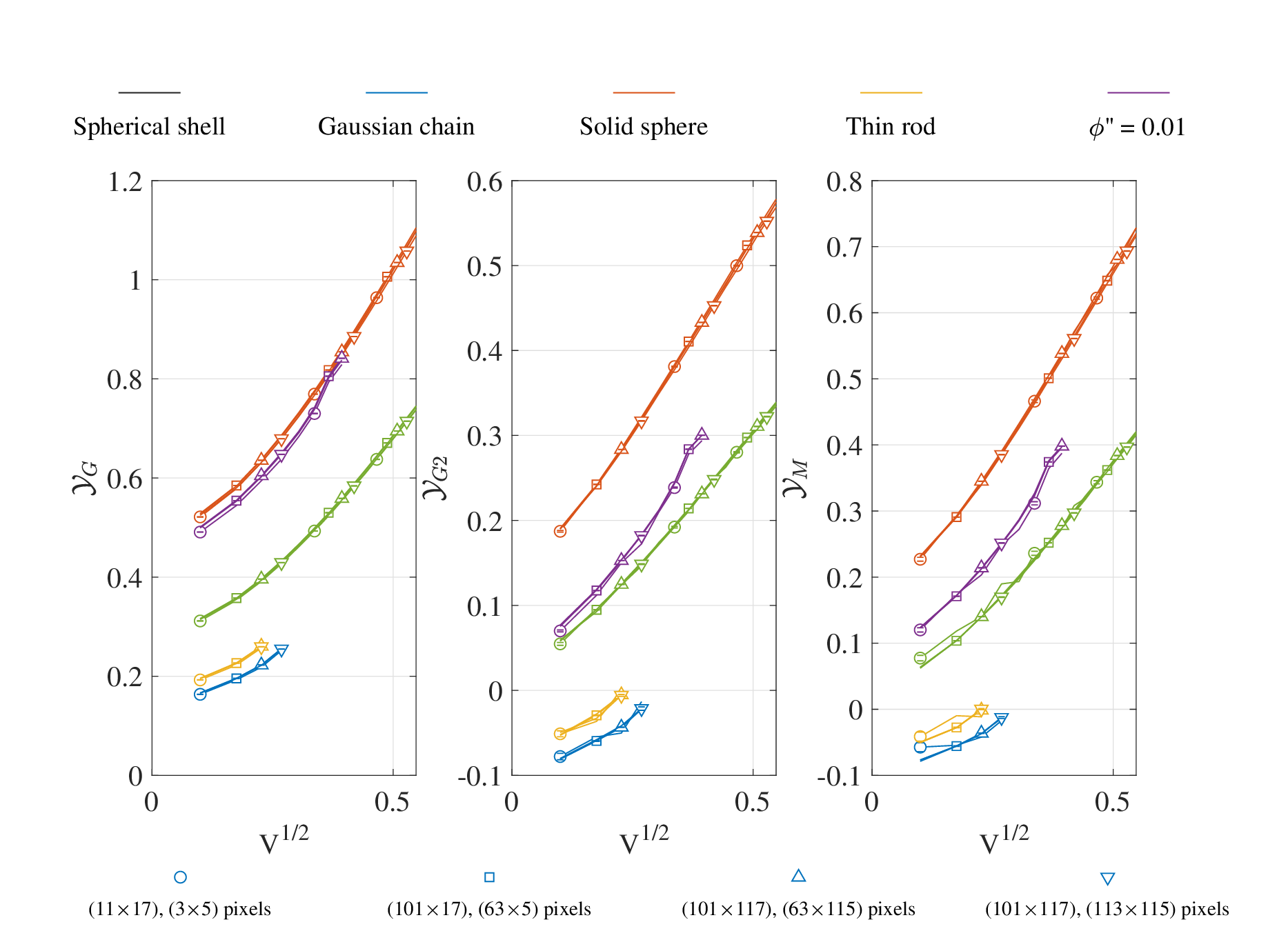}
\caption{Sensitivity to the TENOR-SAXS smearing pixel quartet.}
\label{fig:sensitiVT2}
\end{figure}

In order to check how the observables depend on different parameters, and to find the suitability of the method to be compared to a simulation (calibration), we simulated the scattering for different ensembles, and extracted the observables. We varied either the average radius of gyration, the number of different ensemble members, or the PSF shapes and sizes used for the TENOR-SAXS analysis. The distribution function was also changed (presented in Fig.\,\ref{fig:p_distribution_braid}). The \textit{baseline parameters} were Diamond Light Source B21 beamline characteristics \cite{Cowieson2020_B21} with the scattering ensemble featuring 11 normally distributed members, with $\left<R_g\right>=5\,\text{nm}$. The observables extracted at different conditions are shown in Fig.\,\ref{fig:sensitiVT} - \ref{fig:sensitiVT2}, using the parameters detailed in Tab.\ \ref{tab:beam_simul_param}. The horizontal span of each braid of dashed curves indicates the uncertainty of the deduction of the ensemble's polydispersity given an observation and the form factor. In other words, for a given observable value the uncertainty in $V$ depends on the width of the braid for the variability under examination. 

As expected, we find that the mean radius of gyration, $R_0$, hardly plays a role in the spherical and Guinier form factors (Fig. \ref{fig:sensitiVT}). Figure \ref{fig:sensitiVT3} shows that within the tested range and within the second-order approximation, the native instrumental PSF produced only a small change in the extracted observables, confirming the main claim of the TENOR-SAXS method. The instrumental PSFs checked cover both synchrotron and lab-based typical PSFs. In Fig.~\ref{fig:sensitiVT1} we show that the number of different radii (distribution resolution) does not affect the method's performance.
In contrast to the ensemble and instrumental properties, there is a choice of variable used in the application of the method, the smearing PSF quartet size (as detailed in Sec.~\ref{sec:extraction2D}). Figure \ref{fig:sensitiVT2} shows that within the wide span of smearing kernel sizes, the effect on the observables is indistinct. This means that if using the calibration method (for $\mathcal{Y}_{G2}$ and $\mathcal{Y}_{M}$) one may use any quartet for simulating the ground truth. We did find that in noisy simulations, the smearing PSF is better when it is as large as possible (but within the Guinier regime).

\begin{figure}
\includegraphics[width=1\textwidth,keepaspectratio, trim={0.5cm 1cm 0cm 0cm}, clip]{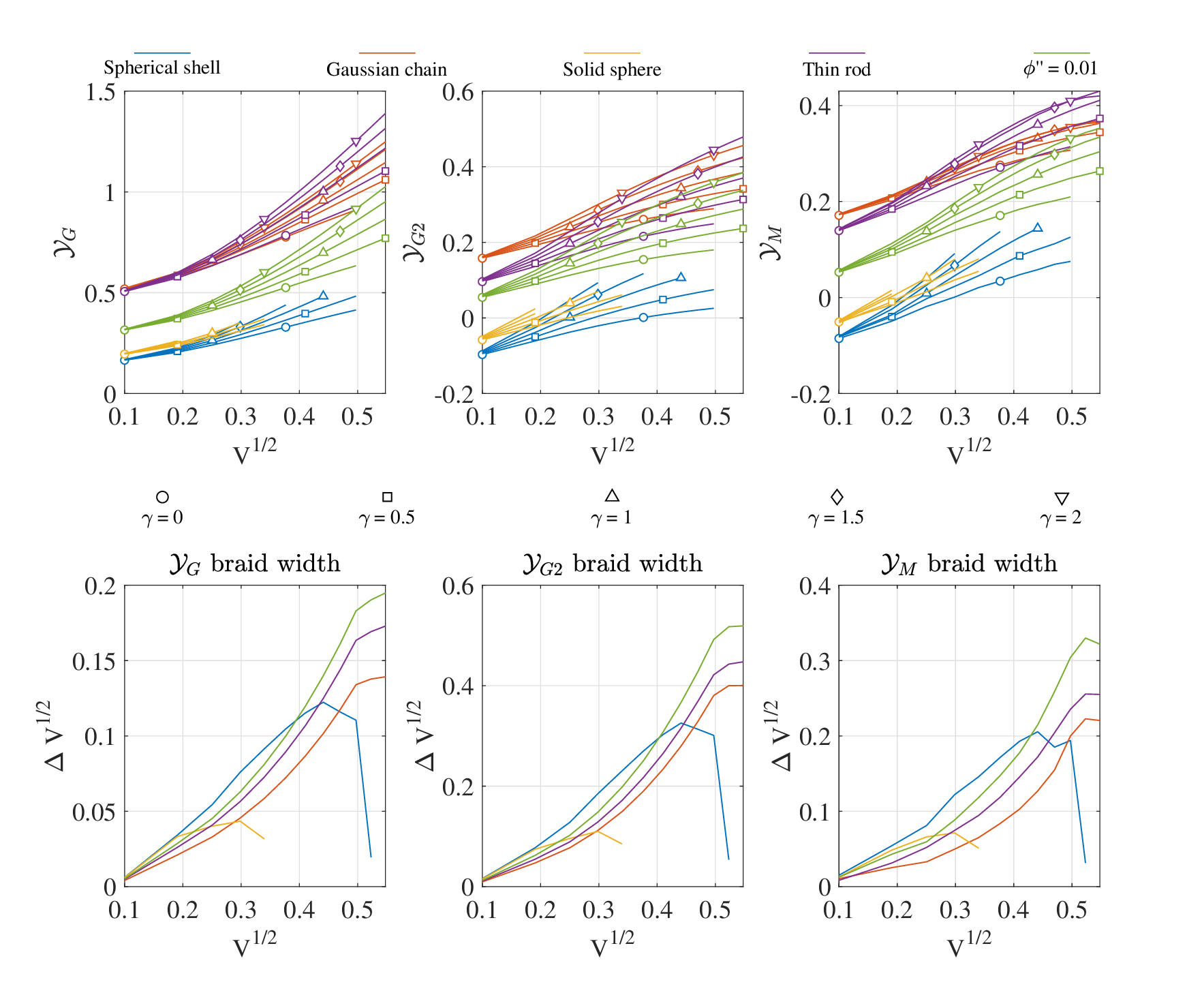}
\caption{Sensitivity to distribution skewness at bimodal distribution. The top panels depict the TENOR-SAXS observables (from left to right, $\mathcal{Y}_G,\mathcal{Y}_{G2},\mathcal{Y}_M$) as derived from noiseless, bimodal distribution having different skewnesses (in the range of $\gamma\in[0,2]$, marker encoded) and form factors (color encoded), at different weighted $V$. The skewness-related $V^{1/2}$ uncertainty is manifested in the width of the braid of equal colored curves (local $\mathcal{Y}$ span divided by mean slope), and is presented in the bottom panels. Simulation parameters were Diamond Light Source instrumentation (Tab.\,\ref{tab:beam_simul_param}) baseline, with $R_0=3 \textrm{ nm}$, and only second order polynomial fit of the log-ratio map. Distributions where the numerical abundance of one of the members was less than 1\% were omitted.}
\label{fig:skewness_sensitivity}
\end{figure}

\textit{Skewness sensitivity:} The sensitivity to specific distribution types was discussed in section \ref{sec:noise_free_sim} and shown in Fig.\ \ref{fig:p_distribution_braid}. In order to quantify the robustness to distribution properties, we studied the skewness dependence of the TENOR-SAXS variance extraction performance. We prepared bimodal distributions with the same $R_0$ and $V$, but with different scattering-weighted skewness $\gamma$ parameter:
\begin{equation}
    \gamma\equiv
    \frac{\left\langle \left(R_g - R_0\right)^3 \right\rangle}
    {R_0^3 V^{3/2}}\qquad.
\end{equation}
We spanned a relatively extreme $\gamma$ span from 0 to 2, omitting cases that require a numerical abundance of less than 1\% and the radii had to be above half the weighted average radius. These constraints were set so that the distribution would be feasible to reproduce in practice and would not be unrealistic. The results, presented in Fig.\ \ref{fig:skewness_sensitivity}, show that for the less sensitive observable $\mathcal{Y}_G$, within the said constraints, the skewness changes the inferred $V^{1/2}$ by up to about $0.1$ for most tested cases (high variance cases start becoming more sensitive). Notice that the high scattering-weight power cases are limited in the span of skewness that supports the constraints, thus these cases are less sensitive than the unweighted cases. We should note that extending the $\gamma$ range to negative values can increase the span, but given the bottom capping of the particle sizes, it seems plausible to assume that the (scattering-weighted) distribution is skewed towards the large particles ($\gamma>0$). 

\textit{Signal strength evaluation}: To obtain a representative estimate of realistic signal-to-noise levels, and to interpret the needed number of collected photons, we consider the Diamond Light Source bioSAXS beamline B21 as an example for a challenging biological case. The beamline delivers $\sim (2-4)\times 10^{12}$~photons per second at the sample \cite{Cowieson2020_B21,Diamond_B21_Website}. 

In absolute units, the macroscopic scattered-photon yield per unit solid angle is $\dot{N}_\Omega \;\approx\; \Phi \, D \, I(q)$, where $\Phi$ is the incident flux (photons $\mathrm{s}^{-1}$) at the sample, $D\approx 1\,\mathrm{mm}$ is the illuminated path length, and $I(q)$ is the absolute scattering intensity expressed in $\mathrm{cm}^{-1}$. 
Typical values of the peak scattering intensity of proteins are $I(0)/c\sim(1-50)\times 10^{-2}~\mathrm{cm}^{2}\,\mathrm{mg}^{-1}$, obtained from the 13.7--669~kDa mass span of \cite{mylonas2007}. Typical protein concentrations are 0.1-10 $\mathrm{mg}\,\mathrm{cm}^{-3}$ \cite{mohammed2024}. Let us take $I(0)\approx1\,\mathrm{cm}^{-1}$, which obtains a scattering yield rate per solid angle of
\[
\dot{N}_{\Omega} \;\approx\; 3\times 10^{11} \ \text{photons\,sec}^{-1}\text{sr}^{-1}.
\]
Converting to the two-dimensional $q$-plane (using the small-angle approximation $q\approx k\theta$) employs the constant Jacobian $d\Omega/d^2q = k^{-2} = \lambda^{2}/(4\pi^{2})$. At $\lambda=0.1\,\mathrm{nm}$ (12.4~keV), matching Tab.\,\ref{tab:beam_simul_param} and the simulations, this factor evaluates to $2.53\times10^{-4}\,\mathrm{sr}\,\mathrm{nm}^{2}$.
Therefore, our estimated forward scattering rate converts to
\[
\dot N_{q^{2}} 
  \approx 7.6\times10^{7}\,
  \mathrm{s^{-1}\,nm^{2}}.
\]
Practically, this means that in such cases the ${\sim}10^{7}\,\mathrm{photons\,nm}^{2}$ accumulated density at which TENOR-SAXS reaches an RMS accuracy of $0.1$ in $V^{1/2}$ (Sec.\,\ref{sec:sim_valid}) is attained well within a second of measurement.


\setcounter{table}{2}
\onecolumn
\begin{table}
\caption{Parameters used in the simulations. The top part lists the resolution based on Diamond Light Source B21 beamline's configuration \cite{Cowieson2020_B21}. The bottom part details the ensemble parameters and the method's variables.} \label{tab:beam_simul_param}
\footnotesize
\begin{tabularx}{\linewidth}{XXXX}
Parameter & Baseline value & Inspected span & Noise benchmark \\
\hline
Wavelength & 0.1 nm & & 0.1 nm \\
Sample-detector distance & 360 cm & & 360 cm \\
Detector pixel size & $70\,\mu\mathrm{m}\times70\,\mu\mathrm{m}$ & & $70\,\mu\mathrm{m}\times70\,\mu\mathrm{m}$ \\
Beam divergence (slit size in pixels) & $3\times15$ & $3\times3$ -- $45\times25$ & $3\times15$ \\
\hline
Mean radius of gyration & $R_0=5\,$nm & 2.5--7\,nm & $\tilde{R}_0=3\,$nm \\
Number of discrete radius values used to represent the ensemble & 11 & 11--61 & 25 \\
Size distribution function & Normal & Boltzmann, Log-normal, Schulz, Triangular, Uniform & Normal \\
 Form factor& $\phi''\in[-0.02,0.06]$&also spherical shell, solid sphere, thin rod, Gaussian chain& Gaussian chain ($\phi''=1/18$)\\
 Smearing PSFs (pixels of $4\sigma$ kernel)& [$125\times123,87\times85$]&[$11\times17,3\times5$] -- [$101\times117,113\times115$]&[$91\times81,117\times127$]\\
 Relative ensemble $R_g$ variance ($V$)& &0--0.3& 12 values, 0--0.3\\
Maximum photon density $N_{q^2}$ (photons\,nm$^2$)& &&$10^{6}$--$10^{9}$\\
 Noise replicates per point& &&30\\
    \end{tabularx}
\end{table}

\twocolumn
\ack{This research was supported by the Pazy Research Foundation (341-2022).
The authors thank Vadim Skoi and Shaked Gefen for helpful discussions.}
\\

\referencelist[saxs2-reviewed-IUCR-abbreviated.bib]
\end{document}